\documentclass[twocolumn]{aastex631}
\usepackage{lipsum}
\usepackage{blindtext}
\usepackage{caption}

\usepackage{amsmath}
\begin{document}

\title{Semblance study of force-field and convection-diffusion solutions with observations of solar magnetic phenomena}

\author{M. Enríquez-Vargas}
\affiliation{Instituto de Astronomía, Universidad Nacional Autónoma de México,
\\ 04510, Ciudad de México, México}
\author{Jorge U. Pérez Romero}
\affiliation{Facultad de Ingeniería, Universidad Nacional Autónoma de México,
\\ 04510, Ciudad de México, México}
\footnote{Computational resources can be found in https://github.com/JorUrie/TRC.git}

\begin{abstract}

We propose a quantitative model using the semblance method to evaluate differences in transport equation solutions and Local Interstellar Spectra (LIS) for Cosmic Ray (CR) modulation. The semblance method quantitatively compares transport solutions—convection-diffusion and force field—and LIS, providing insights into CR behavior.

By integrating these solutions into the atmospheric yield function (Caballero-Lopez \& Moraal 2012), the model identifies a correlation between the modulation factor and sunspots, and an anti-correlation with the mean magnetic field. The force field solution shows stronger semblance values, aligning better with neutron detector data. Among LIS, models by Lagner, Potgieter \& Webber (2003), and Ghelfi et al. (2017) are most consistent, while the Garcia-Munoz et al. (1975) LIS reveals notable variations.

This study underscores the semblance method as a critical tool for comparing CR modulation models, advancing predictive models of cosmic ray behavior influenced by solar and interstellar conditions.

\end{abstract}


\section{Introduction} \label{sec:intro}
To generate the proposed model, data from neutron monitors were utilized and diverse local interstellar spectra were computed. These spectra were introduced into the atmospheric yield function to determine the cosmic ray modulation factor. 

Primary cosmic rays are electrically charged particles and sometimes can reach velocities as high as the speed of light. Primary cosmic radiation is composed of protons, \(\alpha\)-particles (two protons and two neutrons), \(\ ^{-4}\)He and heavy nuclei.

Primary cosmic rays enter the Earth's magnetosphere and interact with the particles in the Earth's atmosphere, those violent collisions produce smaller particles (secondary cosmic rays). This process is repeated multiple times until the particles lose sufficient energy to become elemental particles or are unable to break down due to energy loss. This phenomenon is known as an air shower \citep{Auger}.

High-energy cosmic rays are produced by stochastic collisions within a magnetically turbulent plasma such as a supernova, however there are also low-energy cosmic rays from similar stars to our Sun \citep{sekeido}.

Cosmic rays originating from sources beyond our solar system are detected on Earth's surface, and their detection is affected by solar activity, which is reliant on the solar cycles.
Furthermore, the Sun also generates cosmic rays within its atmosphere. 

The semblance method has been used to compute the desired model. Semblance is a geophysical tool mainly used in exploration seismology, and it can be defined as a correlation in Fourier terms or other transforms between two traces, that is, semblance is a measure to compute the similarity between two traces.

\section{Transport equation} \label{sec:transport equation}

The solutions of the transport equation used in this document are based on Caballero-Lopez and Moraal \citep{Caballero} article.

The Parker's equation is given by:
\begin{equation}
\frac{\partial f}{\partial t} + \frac{1}{4 \pi p^2} \nabla \cdot \vec{S} + \frac{1}{p^2} \frac{\partial}{\partial p} (p^2 \dot{p} ) f = Q 
\label{eq15}  
\end{equation}

where \(S\) is the current differential density:
\begin{equation}
    \vec{S} = 4\pi p^2 (C\vec{V} f - K \cdot \nabla f)
    \label{eq16}
\end{equation}
The equation \ref{eq15} is in terms of the omnidirectional distribution function \(f(r; P; t)\), where \(p\) is the moment of the particle, \(t\) is the time, \(r\) is the radial distance, \(Q\) is the source of cosmic rays and \(V\) the solar wind speed. \(K\) is the diffusion
tensor, that contains a parallel component \(k_ \parallel \), perpendicular \(k_ \perp \) and transverse \(k_T\), and they describe the drifts.
In the equation \ref{eq16}, \(C\) (from the equation \ref{eq17})  is known as the Compton-Getting factor \citep{Gleeson}. This factor was introduced into the transport equation, because the flow is isotropic in the reference system at rest and is not the same in the solar wind system.
\begin{equation}
C = - \frac{1}{3} \frac{\partial ln f}{\partial ln p}
    \label{eq17}
\end{equation}

The adiabatic energy loss is denoted as:
\begin{equation}
    \dot{p} = -\frac{p}{3} \nabla \cdot \vec{V}
    \label{eq18}
\end{equation}

Substituting \ref{eq17} and \ref{eq18} into \ref{eq15} two possible solutions could be obtained:

\small
\begin{equation}
\begin{split}
    \frac{\partial f}{\partial t} + \nabla \cdot ( \underbrace{ C \vec{V} f}_{a} \underbrace{-K \cdot \nabla f}_{b} ) + & 
    \underbrace{ \frac{1}{p^2} \frac{\partial}{\partial p} \left[p^2 \left (\frac{p}{3} \vec{V} \cdot \frac{\nabla f}{f} \right) \right]}_{c} f = Q
    \label{eq19}
\end{split}
\end{equation}

\hypertarget{force-field}{%
\section{\texorpdfstring{\textbf{Force
Field}}{Force Field}}\label{force-field}}
In equation \ref{eq19} the term \(a\) corresponds to the convection, \(b\) to the diffusion and drifts of the cosmic rays, and \(c\) to the changes in energy. To solve the previous equation, different conditions have been considered.
If a steady state, no sources, and negligible energy losses are assumed, the following can be obtained from equation \ref{eq19}:
\begin{equation}
    CVf - K \cdot \nabla f = cnst. = 0
    \label{eq20}
\end{equation}
Equation \ref{eq20} is valid when the term for adiabatic losses in equation \ref{eq21} is smaller than the terms convective and diffusive:
\begin{itemize}
    \item \(V r=k << 1\) valid for high energy cosmic rays, since k
is proportional to energy.
    \item \((1/f) \partial f / \partial r << C/r \) valid for small \(r\), that is, in the internal heliosphere.
\end{itemize}

Considering spherical symmetry, \ref{eq22} is:
\begin{equation}
    CVf - k\frac{\partial f}{\partial r} = 0
    \label{eq21}
\end{equation}

The force field equation in terms of rigidity and knowing that \(P = cp/q\):
\begin{equation}
    \ \frac{\partial f}{\partial r} + \frac{VP}{3k} \frac{\partial f}{\partial P} = 0 
    \label{eq22}
\end{equation}
From:
\begin{equation}
    df(r,P)= \frac{\partial f}{\partial r} dr + \frac{\partial f}{\partial P} dP
    \label{eq23}
\end{equation}

dividing by \(dr\):
\begin{equation}
    \ \frac{df}{dr} = \frac{\partial f}{\partial r} dr + \frac{\partial f}{\partial P} \frac{dp}{dr}
    \label{eq25}
\end{equation}

Considering the equation \ref{eq22}:
\begin{equation}
    \ df = 0
    \label{eq25}
\end{equation}
Equation \ref{eq25} implies that \(f(r, p)\) is constant, equal to
its value at the boundary, along a contour with characteristic equation \(dP/dr = V P/3k\), in the space \((r, P)\).
In the second term of the equation \ref{eq22}, the quantity
\(VP/k\) has units of potential per unit length, that is, force units, this  is why it is called a force field solution.
Considering the diffusion coefficient with the form:
\begin{equation}
    \ k(r,P) = \beta k_1 (r) k_2 (P)
    \label{eq26}
\end{equation}
the solution of the characteristic equation remains as:
\begin{equation}
\begin{split}
    \int_{P}^{P_{b}(r,P)}{\frac{\beta\left( P^{'} \right)k_{2}\left( P^{'} \right)}{P^{'}}dP^{'}} & = \int_{r}^{r_{b}}{\frac{V\left( r^{'} \right)}{3k_{1} (r^{'})}dr'} \equiv \phi(r)
    \label{eq27}
\end{split}
\end{equation}
where \( \phi \) is the force field parameter.
When \(k2 \propto P\) and \( \beta \approx 1\), the solution is reduced to the most commonly used form:
\begin{equation}
    \ \phi = P_b - P
    \label{28}
\end{equation}

\hypertarget{convection-diffusion}{%
\section{\texorpdfstring{\textbf{Convection-Diffusion}}{Convection-Diffusion}}\label{convection-diffusion}}
From equation \ref{eq19}, \(Q\) is not the adiabatic loss in a rest system:
\begin{equation}
    Vf - k\frac{\partial f}{\partial r} = 0
    \label{eq47}
\end{equation}
The outcome is:
\begin{equation}
    f = f_{b}e^{- M}
    \label{eq48}
\end{equation}
Where \(M = \int_{r}^{r_{b}}\frac{Vdr}{k}\) is the modulation function, \(\phi\) is the force field parameter. Its outcome is:
\begin{equation}
    M = \frac{3\phi}{\beta k_{2}}
    \label{eq49}
\end{equation}
Where \(M\) does not have units.

Both solutions are introduced into the atmospheric yield solution, which is necessary to compute the modulation factor. However, LIS is also necessary in order to compute the atmospheric yield solution.

\section{Local Interstellar Spectral} \label{sec:intro}

The Local Interstellar Spectrum is a mathematical model created from observations or simulations and describe the arrival of the Cosmic Rays outside the Heliosphere in energy terms.

The LIS models used in these analyses were Lagner, Potgieter \& Webber LIS in 2003, Burguer \& Potgieter LIS in 2000, Garcia-Munoz, Mason \& Simpson LIS in 1975 and Ghelfi, Barao, Derome \& Maurin LIS in 2017.

\hypertarget{webber-lockwood}{%
\subsection{\texorpdfstring{\textbf{Lagner, Potgieter \& Webber LIS in 2003}}{Lagner, Potgieter \& Webber in 2003}}\label{webber-lockwood}}
This LIS is based on an analytic solution of the Parker equation \citep{Parker}, using Voyager 1, Voyager 2 and Pioneer data at 70 Astronomical Units (AU) close to the Terminal Shock (TS) during the solar minimum between 1987 and 1997.

The modulation developed by Parker included drift effects which are related to the polarity of solar magnetic field:

\begin{equation}
\begin{split}
    \frac{\partial f}{\partial t} = & - \left( V + \left\langle v_{D} \right\rangle \right) \cdot \nabla f + \nabla \cdot \left( K_{S} \cdot \nabla f \right)\\& + \frac{1}{3}(\nabla \cdot V)\frac{\partial f}{\partial InP} + j_{source}
    \label{eq1}
\end{split}
\end{equation}

Where \(j_{source}\) is the local source. In the equation, parallel, perpendicular, and asymmetric drift coefficients were introduced to describe the gradients and drifts of curvature in large-scale Heliospheric Magnetic Field (HMF).

The local sources are not considered and are time-dependent in a spherical coordinate system modeled as a combination of diffusive shock acceleration and drift modulation in two spatial dimensions, neglecting any azimuthal dependence and assuming a symmetrical equatorial field \citep{LangnerAndPotgieter}.
The model offers several CR intensities, which are interpreted as particles of different compositions. When the modulation is computed close to or outside the TS, it shall be computed in rigidity terms. The protons model is the following:
\begin{equation}
    J_{H} = \frac{2.1E_{\frac{k}{n}}^{- 2.8}}{1 + 5.85E_{\frac{k}{n}}^{- 1.2} + 1.18E_{\frac{k}{n}}^{- 2.54}}
    \label{eq2}
\end{equation}
And the Helium model is:
\begin{equation}
    J_{H} = \frac{1.075E_{\frac{k}{n}}^{- 2.8}}{1 + 3.9E_{\frac{k}{n}}^{- 1.09} + 0.90E_{\frac{k}{n}}^{- 2.54}}
    \label{eq3}
\end{equation}
\\
\hypertarget{Burguer \& Potgieter LIS in 2000}{%
\subsection{\texorpdfstring{\textbf{Burguer \& Potgieter LIS in 2000}}{Burguer}}\label{burguer}}
The LIS was compared to ULYSSES data from September 1994 to July 1995 with a Fast Scanning of Latitude (FSL). The Parker equation was analytically solved considering an omnidirectional distribution (two dimensions), the source is located 100 UA, the velocity of the Solar Wind (SW) is \(400\ \frac{km}{s}\) in equatorial plane, and it is increasing to \(800\frac{km}{s}\) in polar zones. The HMF angle sheet is \(15{^\circ}\), which is a good value to fit for solar modulation. The model was computed by Bibber \citep{Bibber}:

\begin{equation}
    f_{IS}(R) = \left\lbrace\begin{array}{c}
    1.9 \times 10^{4}R^{- 2.78}\ \ \ \ \ \ \ \ \ \ \ \ \ \ \ \ \ \ \ \ \ If\ R \geq 7GV \\
    \exp(9.472 - 1.999R - 0.6938R^{2} + 0.2988R^{3} - \\ 0.04714R^{4})\ \ \ \ \ \ \ \ \ \ \ \ \ \ \ \ \ If\ R < 7\ GV 
    \end{array} (\right)
    \label{eq4}
\end{equation}
The model can also be expressed in terms of proton composition:
\begin{equation}
    J_{H} = \frac{1.9 \times 10^{4}P^{- 2.78}}{1 + 0.4866P^{- 2.51}}
    \label{eq5}
\end{equation}

\begin{equation}
    J_{He} = \frac{\left( 3.8 \times 10^{4}P^{- 2.78} \right)}{\left( 1 + 0.9732P^{- 2.51} \right)}
    \label{eq6}
\end{equation}
\hypertarget{Garcia-Munoz}{%
\subsection{\texorpdfstring{\textbf{Garcia-Munoz, Mason and Simpson LIS in 1975}}{Garcia-Munoz}}\label{garcia-munoz}}
To compute this LIS, IMP-5, IMP-7, and IMP-8 data were consulted, because the satellites can capture several particles and their energy range is the widest. The selected data" were electrons \textgreater{} \(100MeV\) with quiet-time measurements.

The Parker's equation was analytically solved considering a symmetric spherical coordinate system in steady state and drawing on Fisk's method \citep{Fisk}.
The LIS is the following:
\begin{equation}
    j = A(T +  B exp( - CT))^{\gamma}
    \label{eq7}
\end{equation}
Where \(j\) is particle flux, \(T\) is kinetic energy and parameters are shown in Table \ref{table1}:
\begin{table}[!ht]
\begin{center}
\begin{tabular}{ |c|c|c|c|c| } 
\hline
Species & A & B & C & $\gamma$ \\ 
\hline
H & $9.9\times 10^8$ & 780 & $2.5\times 10^-4$ & 2.65 \\ 
$ ^4 He$ & $1.4\times 10^8$ & 660 & $1.4\times 10^-4$ & 2.77\\ 
C & $1.8\times 10^6$ & 620 & $5.2\times 10^-4$ & 2.68\\ 
\hline
\end{tabular}
\end{center}
\caption{\label{table1}Equation 7 parameters \cite{Garcia-Munoz}}
\end{table}

\hypertarget{Maurin}{%
\subsection{\texorpdfstring{\textbf{Ghelfi, Barao, Derome and Maurin LIS in 2017}}{Maurin}}\label{maurin}}
The data used were AMS-01 and AMS-02, BESS-Polar and PAMELA, because they were recent (at that time), but Voyager 1 data was also considered.

First, the author proposes the simplest modulation model to link unmodulated (IS) to modulated (TOA) quantities, which is a force field approximation  \cite{Ghelfi}:
\begin{equation}
    \frac{E^{TOA}}{A} = \frac{E^{IS}}{A} - \frac{|Z|}{A}\phi
    \label{8}
\end{equation}
\begin{equation}
    J^{TOA}\left( E^{TOA} \right) = \left( \frac{p^{TOA}}{p^{IS}} \right)^{2} \times J^{IS}\left( E^{IS} \right)
    \label{9}
\end{equation}

Where \(E\) is total energy, \(p\) is the momentum, and \(J \equiv \frac{dJ}{dE_{\frac{k}{n}}}\) is the differential flux per kinetic energy per nucleon \(E_{\frac{k}{n}}\).

An analysis \(\chi^{2}\) was necessary to fix the TOA flux with all the species \(N_{j}(i)\) in this \(t_{j}\), over all possible energy \(E_{k}(i,\ j)\) and is given by:
\begin{equation}
    \chi^{2} = \sum_{t_{i}}^{}{\sum_{N_{j(i)}}^{}{\sum_{E_{k(i,j)}}^{}\frac{\left( J^{TOA}\left( J_{j}^{IS},\ \phi_{j},E_{k} \right) - data_{ijk} \right)^{2}}{\sigma_{ijk}}}}
    \label{10}
\end{equation}

Where IS parameters are free. For the previously mentioned cases, power laws in total energy \citep{O'Neill} or rigidity \citep{Shikaze}.
TOA data for H and He are old and have more inconsistencies than new data, thus, a \(\chi^{2}\) analysis was applied to minimize errors. Therefore, the MINUIT minimization package was used \citep{James:2296388} from Root CERN libraries \cite{Sartini}.
Markov Chain Monte Carlo analysis (MCMC) was implemented from the GreAT package \citep{Putze} to determine the correlation between credible intervals for \(\phi\) and IS flux. MCMC analysis is based on Bayes' theorem.
Finally, all the obtained data is represented as an interpolation of logarithmic-polynomial functions, as shown in equation \ref{eq11}.
\begin{equation}
\begin{split}
    \log_{10}\left( J_{IS} \right) = \left\{ \begin{array}{ll}
    \sum_{i = 0}^{12}c_{i}\left( \frac{\log_{10}\left( E_{\frac{k}{n}} \right)}{\log_{10}\left( 800_{\frac{GeV}{n}} \right)} \right)^{i} & \text{if } E_{\frac{k}{n}} < 800\frac{GeV}{n} \\[10pt]
    C_{0} - C_{1}\left( \frac{\log_{10}\left( E_{\frac{k}{n}} \right)}{\log_{10}\left( 800_{\frac{GeV}{n}} \right)} \right) & \text{if } E_{\frac{k}{n}} \geq 800\frac{GeV}{n} \\
    \end{array} \right.
    \label{eq11}
\end{split}
\end{equation}

In this paper, just four LIS were applied, however more models were also considered, and it is worthwhile to mention them.

\hypertarget{Della Torre}{%
\subsection{\texorpdfstring{\textbf{Boschini, Della Torre, Gervasi and more LIS in 2018}}{Della Torre}}\label{della torre}}
For the calculation of this LIS the GALPROP and HelMod programs were used to simulate CR data, but real data was required, so Voyager 1, BESS, PAMELA, AMS-01 and AMS-02  data were consulted.

The GALPROP code uses astronomical, particle physics and nuclear information
to predict CR flux, X-ray, synchrotron emissions and their polarization \citep{Strong_2007}.
Heliosphere propagation was computed by GALPROP and thanks to the MCMC analysis, the data became more realistic. Afterward, the information was introduced into the HelMod program to obtain modulated data.

Previous data was compared with real information from the satellites with different energy content. The used information was a combination of Voyager 1, AMS-02, CREAM-I and ATIC-02, but CREAM-I and AMS-02, because they had better adjustments and offered minimal errors. To compute the analytical LIS, MCMC  analysis from Eureqa\footnote{http://www.nutonian.com/products/eureqa/} was used, the outcome is given:
\begin{equation}
    F(R) \times R^{2.7} = \left\{ \begin{array}{c}
    \sum_{i = 0}^{5}{a_{i}R^{i}}\ \ \ \ \ \ \ \ \ \ \ \ \ \ \ \ R \leq 1GV \\
    b + \frac{c}{R} + \frac{\left( d_{1} \right)}{d_{2} + R} + \frac{\left( e_{1} \right)}{e_{2} + R} + \frac{\left( f_{1} \right)}{f_{2} + R} + \\ gR\ \ \ \ \ \ \ R \geq 1GV \\
    \end{array} \right.\ 
    \label{12}
\end{equation}
Where the parameters are showing in the table 2.

\begin{figure}[h]
    \centering
    \includegraphics[width=8cm]{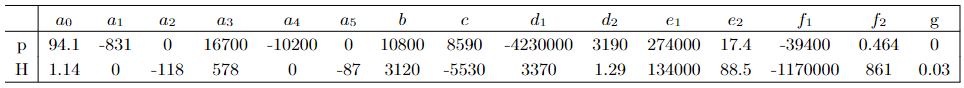}
    \captionsetup{labelformat=empty}
    \caption{Table 2. Equation 12 parameters \citep{DellaTorre}}
\end{figure}

\hypertarget{Vos \& Marius}{%
\subsection{\texorpdfstring{\textbf{Vos \& Potgieter LIS in 2015}}{Vos \& Marius}}\label{VosAndMarius}}
The LIS was computed using observations from PAMELA data during a solar minimum from 2006 to 2009. This LIS was computed from the Parker equation \citep{Parker}, which was solved as a ray distribution function in  rigidity terms, time, position in 3D, in heliocentric spherical coordinates where the polar angle is equal to \(90 ^{\circ}\). All short-term modulation effects were neglected (solar minimum conditions). The average particle drift velocity is caused by gradients and curvature in Heliospheric Magnetic Field (HMF), there is a symmetric diffusion tensor and adiabatic energy changes except in the heliosheath \citep{VosII}. 

The LIS is the following \citep{Vos}:
\begin{equation}
    j_{LIS} = 2.70\left( \frac{E^{1.12}}{\beta^{2}} \right)\left( \frac{E + 0.67}{1.67} \right)^{- 3.93}
    \label{eq13}
\end{equation}
Where \(E\) is kinetic energy, \(\beta = \frac{v}{c}\) particle velocity (like light velocity), \(j_{LIS} = P^{2}f\).

\hypertarget{Moskalenko}{%
\subsection{\texorpdfstring{\textbf{Moskalenko, Strong, Ormes and Potgieter LIS in 2002}}{Moskalenko}}\label{Moskalenko}}
The cosmic ray data was simulated using the DTUNUC Monte Carlo code  \citep{Ferrari_1996} , and additional data was collected from two previous works \citep{TanandNg}.
The Parker equation \citep{Parker} was solved through Crank-Nicholson
numerical solution \citep{crank_nicolson_1947} and according to the author: a more realistic model was sought. Thus, the leaky box code (weighted-slab) was applied. GALPROP was also used to compute 3D cosmic rays.

The Helium LIS was approximated using a force field solution due to the low-energy approximation, and the \(\phi\) potential modulation was selected using the CLIMAX neutron monitor.

To ensure that the spectrum is similar to the simulated values, a power law dependent on kinetic energy based on the convection-diffusion solution has been approximated, which cannot only be described by a single function.

Finally, the LIS was subjected to a \(\chi_{n}^{2}\) function to observe the
quality of the adjusted data for each individual measurement
\citep{Moskalenko_2002} as shown in equation \ref{eq14}:
\begin{equation}
\begin{split}
    J_{H} = \left\{ \begin{array}{l}
    \exp\left( 4.64 - 0.08\left( \log\left( E_{\frac{k}{n}} \right) \right)^{2} -  2.91\sqrt{E_{\frac{k}{n}}} \right),  \\ \ \ \ \ \ \ \ \ \ \ \ \ \ \ \ \ \ \ \ \ \ \ \ \ \ \ \ \ \ E_{\frac{k}{n}} \leq 1GeV \\
    \exp\left( 3.22 - 2.86\log\left( E_{\frac{k}{n}} \right) - \frac{1.5}{E_{\frac{k}{n}}}\  \right), \\ \ \ \ \ \ \ \ \ \ \ \ \ \ \ \ \ \ \ \ \ \ \ \ \ \ \ \ \ \ \ E_{\frac{k}{n}} > 1GeV \\
    \end{array} \right.\ 
\label{eq14}
\end{split}
\end{equation}

The previously mentioned LIS are shown in the figure \ref{LIS} using the rigidity-purpose parameters defined by each author:
\begin{figure}[htb]
    \centering
    \includegraphics[width=0.5\textwidth]{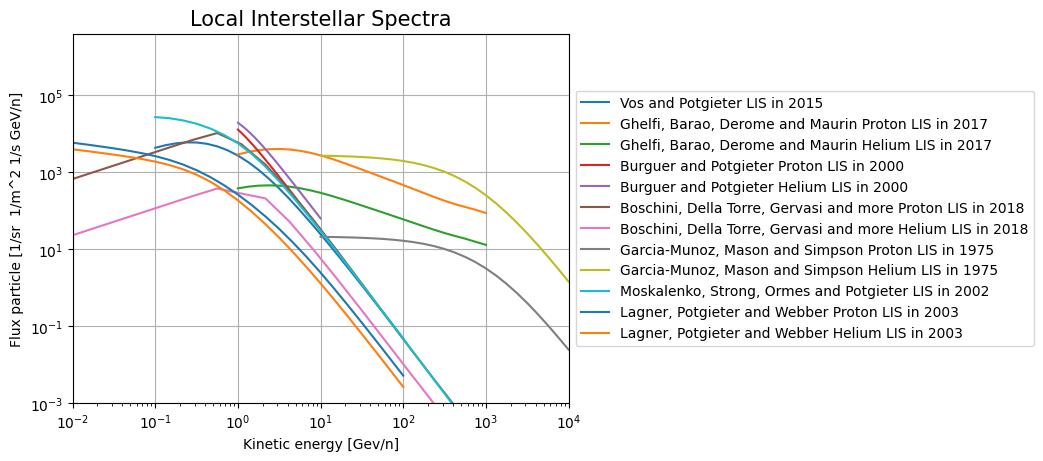}
    \caption{Local Interstellar Spectral purpose by \citep{Vos}, \citep{Ghelfi}, \citep{Burguer}, \citep{DellaTorre}, \citep{Garcia-Munoz}, \citep{Moskalenko_2002} and \citep{LangnerAndPotgieter}}
    \label{LIS}
\end{figure}

\hypertarget{semblance}{%
\section{\texorpdfstring{\textbf{Semblance}}{Semblance}}\label{semblance}}
The semblance is the cross-relation between two traces, the computed value is interpreted as the similarity of two points contained in the traces within the Fourier domain. However, in this analysis, the Continuous Wavelet Transform (CWT) was used which represents a wave in two dimensions: data number and wavelength.

CWT is given:
\begin{equation}
    CWT(u,s) = \int_{- \infty}^{\infty}{h(t)\frac{1}{|s|^{0.5}}\Psi^{*}\left( \frac{t - u}{s} \right)dt}
    \label{eq50}
\end{equation}

Where \(s\) is the scale, \(u\) is the displacement, \(\Psi\) is the mother wavelet , and \({}^{*}\) means complex conjugate and \(t\) is the coordinate in time \citep{Cooper} Equation \ref{eq50} can also be expressed in spatial terms \citep{Teolis}:
\begin{equation}
    \Psi(x) = \frac{1}{\pi f_{b}}e^{2\pi if_{c}x}e^{- \frac{x^{2}}{f_{b}}}
    \label{eq51}
\end{equation}
The conventional semblance is a normalized coherence measure, first computed in 1969 \citep{Taner},later, Neidell and Taner \citep{Neidell} found that the coherence and semblance were two different operators. Thus, semblance was expressed as shown in the equation \ref{eq52}:
\begin{equation}
    S_{NT}\ \lbrack i\rbrack = \frac{\sum_{j = i - M}^{i + M}\left( \sum_{k = 0}^{N - 1}{q\lbrack j,k\rbrack} \right)^{2}}{N\sum_{j = i - M}^{i + M}{\sum_{k = 0}^{N - 1}{q\lbrack j.k\rbrack^{2}}}}
    \label{eq52}
\end{equation}
Where \(i,\ j\) are time sample indices, \(k\) is a trace number, \(q\lbrack j,\ k\rbrack\) is the trace amplitude at time index \(j\) a trace number \(k\) of the NMO-corrected gather.

To reduce and smooth the decays a boxcar filter is often applied, that can be written as \citep{Luo}:
\begin{equation}
    S_{C}\ \lbrack i\rbrack = \frac{\sum_{j}^{}{h\lbrack i - j\rbrack\left( \sum_{k}^{}{q\lbrack j,k\rbrack} \right)^{2}}}{N\sum_{j}^{}{h\lbrack i - j\rbrack\sum_{k}^{}{q\lbrack j.k\rbrack^{2}}}}
    \label{eq53}
\end{equation}
There is an alternative expression for the conventional semblance, which can be expressed as a normalized correlation coefficient. However, first, a trace \(r\lbrack j\rbrack\) should be defined as a summation over the trace number:
\begin{equation}
    r\lbrack j\rbrack \equiv \sum_{k}^{}{q\lbrack j,k\rbrack}
    \label{eq54}
\end{equation}
Therefore:
\begin{equation}
    C_{rq}\lbrack i\rbrack \equiv \sum_{j}^{}{h\lbrack i - j\rbrack\sum_{k}^{}{r\lbrack j\rbrack q\lbrack j,k\rbrack}}
    \label{eq55}
\end{equation}
\begin{equation}
    C_{rr}\lbrack i\rbrack \equiv \sum_{j}^{}{h\lbrack i - j\rbrack\sum_{k}^{}{r\lbrack j\rbrack^{2}}}
    \label{eq56}
\end{equation}
\begin{equation}
    C_{qq}\lbrack i\rbrack \equiv \sum_{j}^{}{h\lbrack i - j\rbrack\sum_{k}^{}{q\lbrack j,k\rbrack^{2}}}
    \label{eq57}
\end{equation}
Now, the three equations \ref{eq55}, \ref{eq56} and \ref{eq57} can be expressed as a conventional semblance \citep{Luo}:
\begin{equation}
    S_{C}\lbrack i\rbrack = \frac{C_{rq}\lbrack i\rbrack^{2}}{C_{rr}\lbrack i\rbrack C_{qq}\lbrack i\rbrack}
    \label{58}
\end{equation}

\section{\texorpdfstring{\textbf{Count rates from ground-based detectors}}{Count rates from ground-based detectors}}\label{count-rates-from-ground-based-detectors}

The neutrons detected at ground level are the product of atmospheric showers caused by the collision of primary particles with molecules in the atmosphere. The neutron count is given by the following equation:

\begin{equation}
\begin{split}
    N\left( P_c,\ \ x,\ \ t \right)  = & \ \int_{P_{c}}^{\infty}\left( \frac{- dN}{dP} \right)dP  = \\ & \sum_{i\ }^{}{\int_{P_{c}}^{\infty}{S_{i}(P,\ x)j_{i}(P,t)dP}}
    \label{59}
\end{split}
\end{equation}
Where \(j_{i}(P,t)\) is the spectrum of the primary species above the atmosphere, and \(S_{i}(P,x)\) is the atmospheric yield function due to this species. \(P_{c}\) is the cut-off rigidity, which is the necessary minimum energy for a particle to enter the Earth's magnetosphere.
The quantity \(dN=dP\) is the differential counting rate of the
instrument inside the atmosphere.
The yield function used in this analysis is the following \citep{Caballero-Lopez}:
\begin{equation}
    S_H = \frac{-(dN/dP)}{J_H(P)+1.584F(P)j_{He}}
    \label{eq58}
\end{equation}
where \(F(P)\) is the ratio between the yield functions of
\(He\) and \(H\) as reported by \citep{Dorman}:
\begin{equation}
    \ F(P) = F_0 (P^a _0 + P^a)^{(\gamma_1-\gamma_2/a)} P\gamma^2
    \label{eq59}
\end{equation}
the values in the equation are: \(F_0 = 2.0\), \( \gamma_1 = 0 \), \( \gamma_2 =10\), \(a=1.4\) and \(P_0 = 0.45\).

The equation \ref{eq58} allows computing a modulation factor  and will be used to obtain the semblance. The \(J_H\) and \(J_{He}\) terms in the equation \ref{eq58} are the transport equation solutions (convection-diffusion or force field), which are in terms of a LIS previously mentioned. A Python code has been programmed and can be found at https://github.com/JorUrie/TRC.git

\hypertarget{first-analysis}{%
\section{\texorpdfstring{\textbf{First
Analysis}}{First Analysis}}\label{first-analysis}}

\begin{figure}[!ht]
    \centering
     \includegraphics[width=0.35\textwidth]{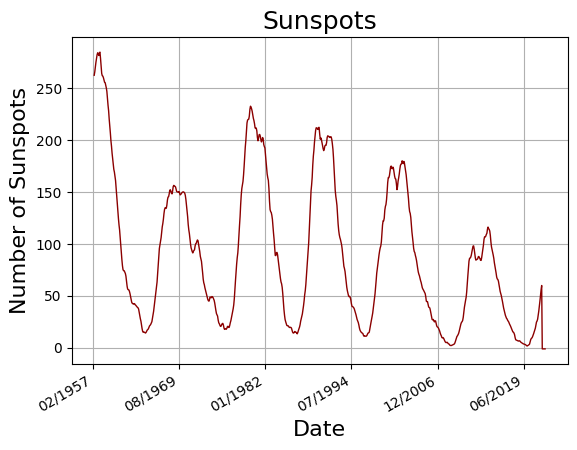}
    \caption{Sunspots data obtained from Sunspot index and Long-term Solar Observations.}
    \label{fig1}
\end{figure}

\begin{figure}[ht!]
    \centering
    \includegraphics[width=0.35\textwidth]{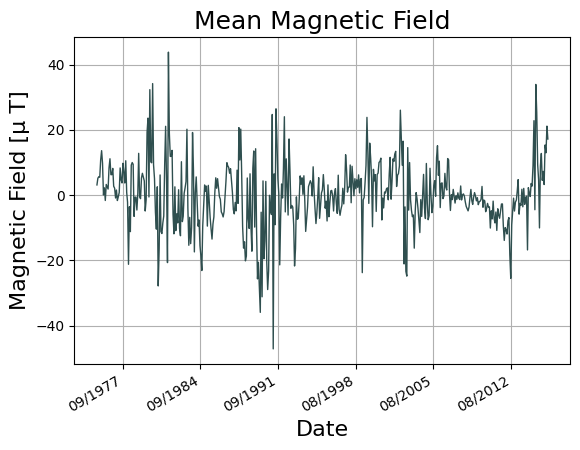}
    \caption{Solar Magnetic Field date obtained from The Wilcox Solar Observatory.}
    \label{fig2}
\end{figure}

The first analysis consists of three stations located in different magnetospheric latitudes (Table \ref{FACR}) and and computing their modulation factor, then semblance between the modulation factor and sunspots or mean solar magnetic field will be computed. The data were downloaded from the Neutron Monitor DataBase (NMDB)\footnote{\href{https://www.nmdb.eu/}{NMDB}}, sunspot data from Sunspot Number\footnote{\href{https://www.sidc.be/silso/datafiles}{Sunspot Number \textbar{} SILSO (sidc.be)}} (Figure \ref{fig1}) and mean solar magnetic field data from The Wilcox Solar Observatory\footnote{\href{http://wso.stanford.edu/}{WSO -The Wilcox Solar Observatory (stanford.edu)}} (Figure \ref{fig2}).
  
\begin{table}[!ht]
\begin{tabular}{|c|c|}
\hline
\textbf{Neutron Monitor} & \textbf{Cut-off Rigidity {[}GeV{]}} \\ \hline
\textbf{Antarctic (SNAE)}            & 0.73                                       \\ \hline
\textbf{Finland (OULU)}            & 0.81                                       \\ \hline
\textbf{Kazakhstan (AATB)}            & 5.90                                       \\ \hline
\end{tabular}
\caption{Cut-off Rigidity from the stations used in the first analysis}
\label{FACR}
\end{table}

The Figure \ref{fig1} shows the number of sunspots recorded, which are related to solar maximums and minimums. The data covers from May 1957 to July 2022. In the same figure, solar cycles can be observed, in other words, solar maximums belong approximately to the years 1958, 1970, 1980, 1990, 2014 and the next solar maximum could be in 2024. 
Solar minimums belong approximately to the years 1965, 1976, 1986, 1997 and 2010. With the previous information, 11-year cycle can be perceived. In 1958 was the largest recorded maximum, while the smallest maximum was in 1970. Conversely, minimums have similar values and there are not many differences. 

The mean magnetic field information belongs to the dates from May 1975 to December 2015. The information has cusps and appears not to have significant changes, but in September 1981 was the biggest registered maximum, while in March 1991 was the smallest registered value. If a line is traced along the graphic following a “regression”, maximums and minimums can be identified, they would be mostly inversely proportionate to sunspots.

\begin{figure}[htb!]
    \centering
    \includegraphics[width=0.35\textwidth]{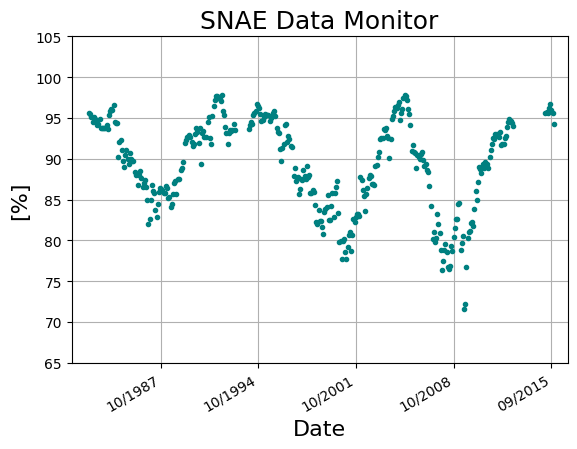}
    \caption{SNAE data obtained from NMDB.}
    \label{fig3}
\end{figure}

\begin{figure}[htb!]
    \centering
    \includegraphics[width=0.35\textwidth]{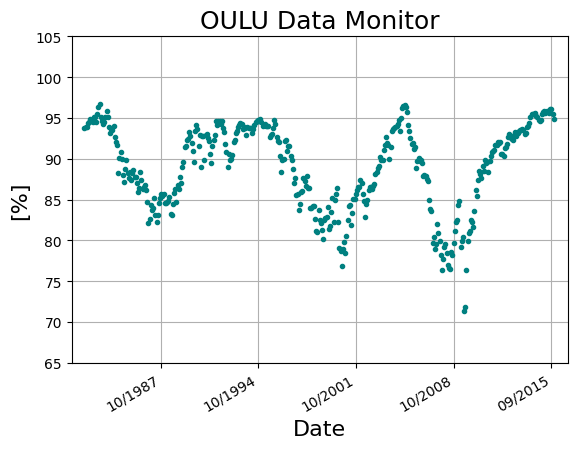}
    \caption{OULU data obtained from NMDB.}
    \label{fig4}
\end{figure}

\begin{figure}[htb!]
    \centering
    \includegraphics[width=0.35
    \textwidth]{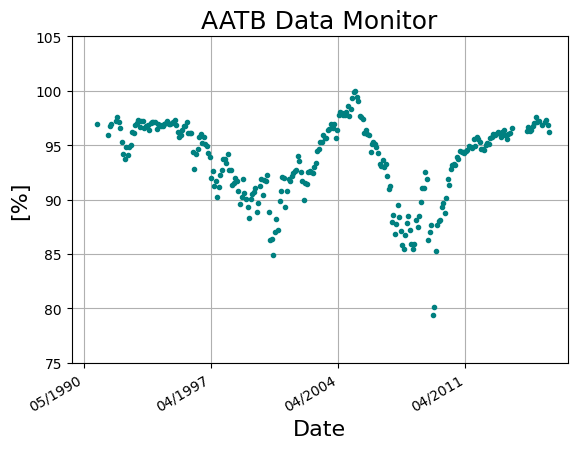}
    \caption{AATB data obtained from NMDB.}
    \label{fig5}
\end{figure}

The selected stations were SNAE (south), OULU (north) and AATB (magnetic equator). They were deemed appropriate based on their respective latitudes in the Earth (Figure \ref{StationsWorld}).
SNAE, OULU and AATB data are shown in the figures \ref{fig3}, \ref{fig4} and \ref{fig5}. The data were downloaded and normalized to the highest contained value.

The previously mentioned Figures \ref{fig3}, \ref{fig4} and \ref{fig5} are normalized data, because the data is obtained from the neutron monitors on the Earth. However, the particles have already been modified by modulation phenomena and they have also lost energy (secondary cosmic rays), for that reason a modulation factor is necessary to compute.

Non-normalized data have been introduced into the atmospheric yield function, which requires both a LIS and a transport equation solution (convection-diffusion or force field) to compute the modulation factor. 

\begin{figure}[htp!]
    \centering    
    \includegraphics[width=0.35 \textwidth]{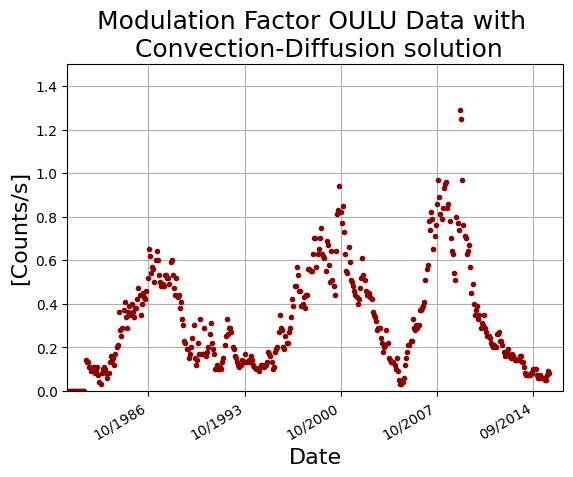}
    \caption{OULU station. Modulation factor with Convection-Diffusion solution using Lagner, Potgieter and Webber LIS in 2003.}
    \label{fig6}
\end{figure}

\begin{figure}[htbp!]
    \centering   
    \includegraphics[width=0.35 \textwidth]{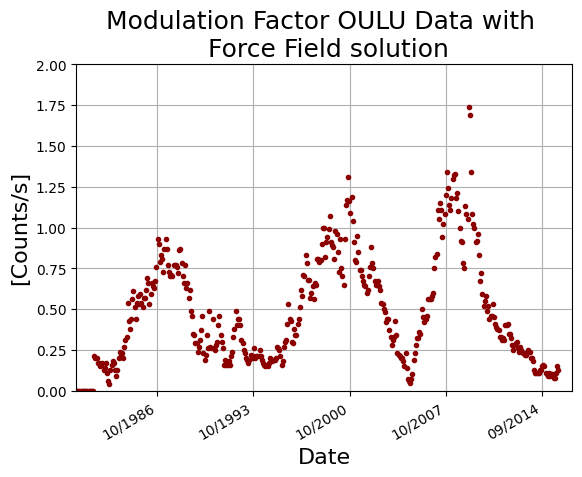}
    \caption{OULU station. Modulation factor with Force Field solution using Lagner, Potgieter and Webber LIS in 2003.}
    \label{fig7}
\end{figure}

\begin{figure}[ht!]
    \centering    
    \includegraphics[width=0.35\textwidth]{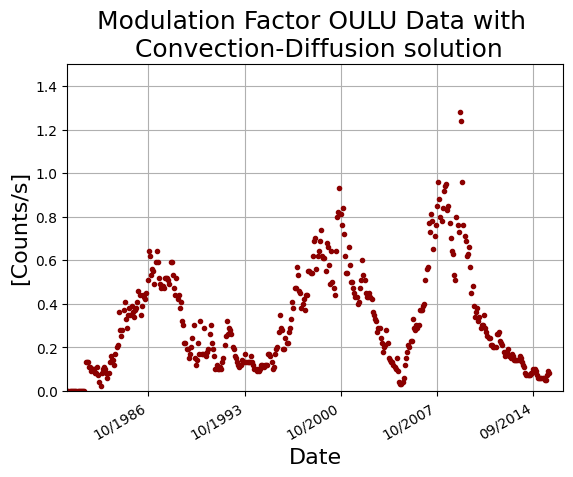}
    \caption{OULU station. Modulation factor with Convection-Diffusion solution using Burguer and Potgieter LIS in 2000.}
    \label{fig8}
\end{figure}

\begin{figure}[ht!]
    \centering   
    \includegraphics[width=0.35\textwidth]{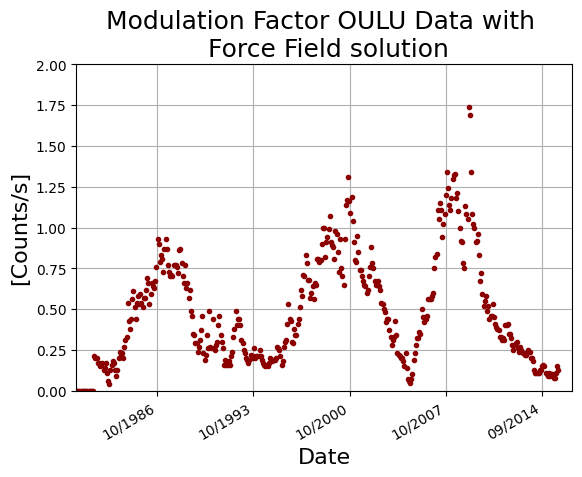}
    \caption{OULU station. Modulation factor with Force Field solution using Burguer and Potgieter LIS in 2000.}
    \label{fig9}
\end{figure}

\begin{figure}[ht!]
    \centering    
    \includegraphics[width=0.35\textwidth]{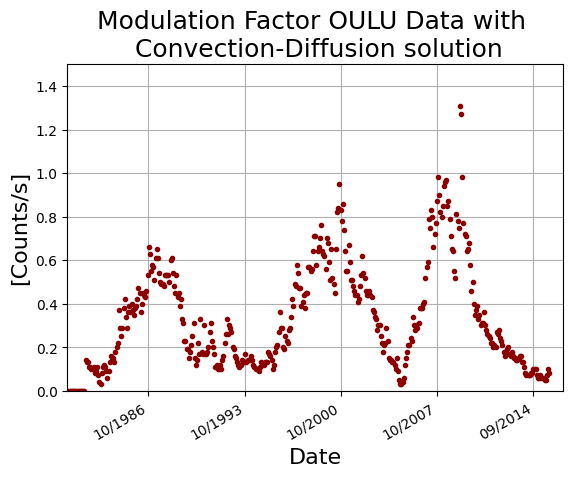}
    \caption{OULU station. Modulation factor with Convection-Diffusion solution using Garcia-Munoz, Mason and Simpson LIS in 1975.}
    \label{fig10}
\end{figure}

\begin{figure}[ht!]
    \centering   
    \includegraphics[width=0.35\textwidth]{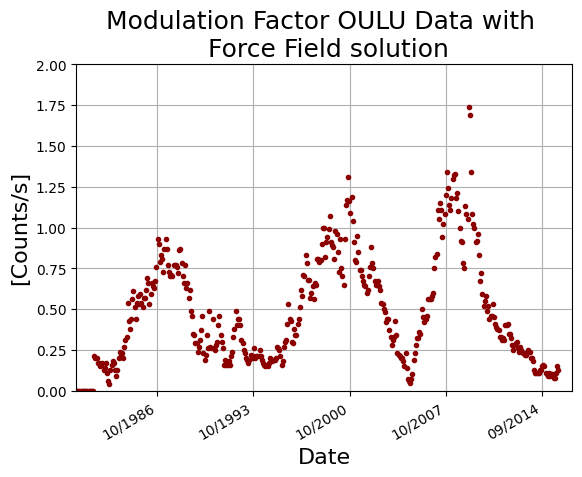}
    \caption{OULU station. Modulation factor with Force Field solution using Garcia-Munoz, Mason and Simpson LIS in 1975.}
    \label{fig11}
\end{figure}

\begin{figure}[ht!]
    \centering    
    \includegraphics[width=0.35\textwidth]{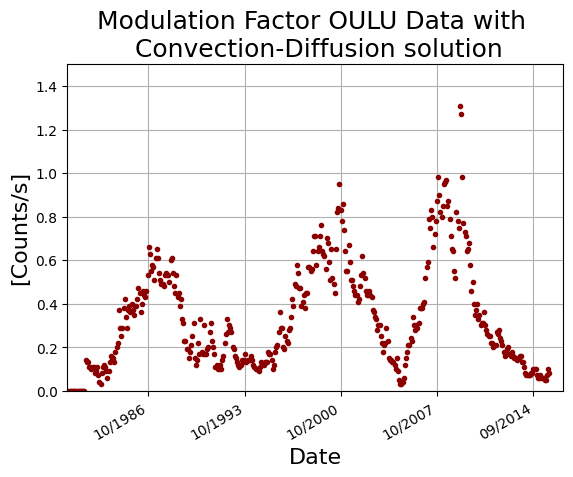}
    \caption{OULU station. Modulation factor with Convection-Diffusion solution using Ghelfi, Barao, Derome and Maurin LIS in 2017.}
    \label{fig12}
\end{figure}

\begin{figure}[ht!]
    \centering   
    \includegraphics[width=0.35\textwidth]{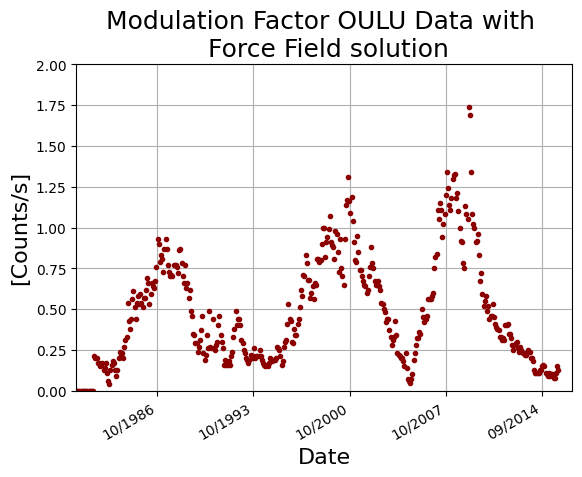}
    \caption{OULU station. Modulation factor with Force Field solution using Ghelfi, Barao, Derome and Maurin LIS in 2017.}
    \label{fig13}
\end{figure}

The figures from \ref{fig6} to \ref{fig13} are the modulation factors for the same station using every previously mentioned LIS and a transport equation solution. Their unities are \(Counts/s\) and \(Date\), because the modulation factor is the amount of particles before they enter the Earth's atmosphere and interact with the particles in a given date. Modulation factor would not be necessary if the satellites would take desired measurements, but the technology and the conditions do not allow it. 

The modulation factors were computed for all selected stations. Those have a similar pattern for each LIS and are also similar to the downloaded data, for example, in Figures \ref{fig4} and \ref{fig6}, high values can be observed, in 1987 near to 2004 and 2011, These patterns also occur in Figures \ref{fig6} to \ref{fig13}. In fact, in Figures \ref{fig6}, \ref{fig8}, \ref{fig10} and \ref{fig12} between values \(1.2\) and \(1.4\) \(Counts/s\), two points are always present and are also the highest values. For the figures \ref{fig7}, \ref{fig9}, \ref{fig11} and \ref{fig13}, the same pattern is observed and the two highest points are also present between \(1.5\) and \(1.75\) \(Counts/s\). The main difference among modulation factors using convection-diffusion and force field solutions is the amplitude.

From Figures  \ref{fig6} to \ref{fig13} also looks like the figure \ref{fig1}, because more extragalactic cosmic rays are introduced into the Earth as a result of low solar modulation.

For the three stations selected according to their latitude, the semblance for the modulation factor and sunspots using several LIS are shown. However, only the results for the OULU station are presented here; the remaining models can be found in the appendix.

\begin{figure}
    \centering
    \includegraphics[width=0.5\textwidth]{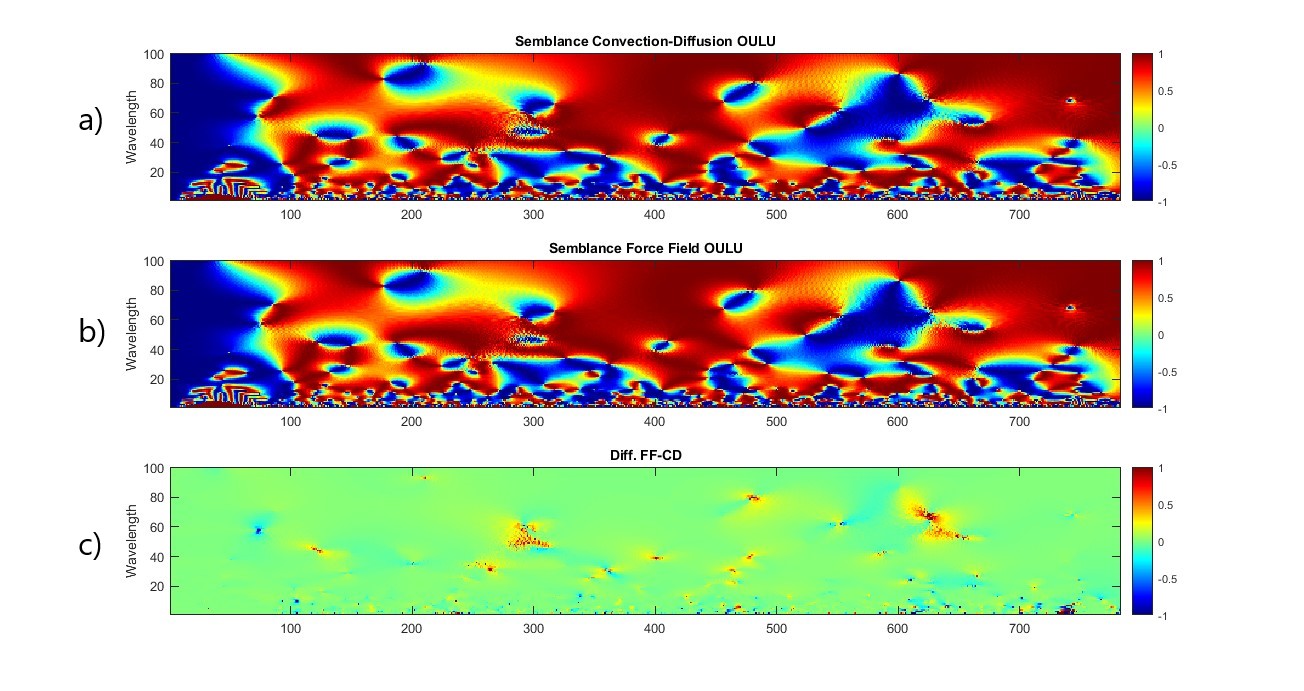}
    \caption{OULU Station a) Semblance between Modulation Factor using Convection-Diffusion with Lagner, Potgieter and Webber LIS in 2003 vs Sunspot data. b) Semblance between Modulation Factor using Force Field with Lagner, Potgieter and Webber LIS in 2003 vs Sunspot data. c) Difference between the two previous semblances.}
    \label{sem2}
\end{figure}

\begin{figure}
    \centering
    \includegraphics[width=0.5\textwidth]{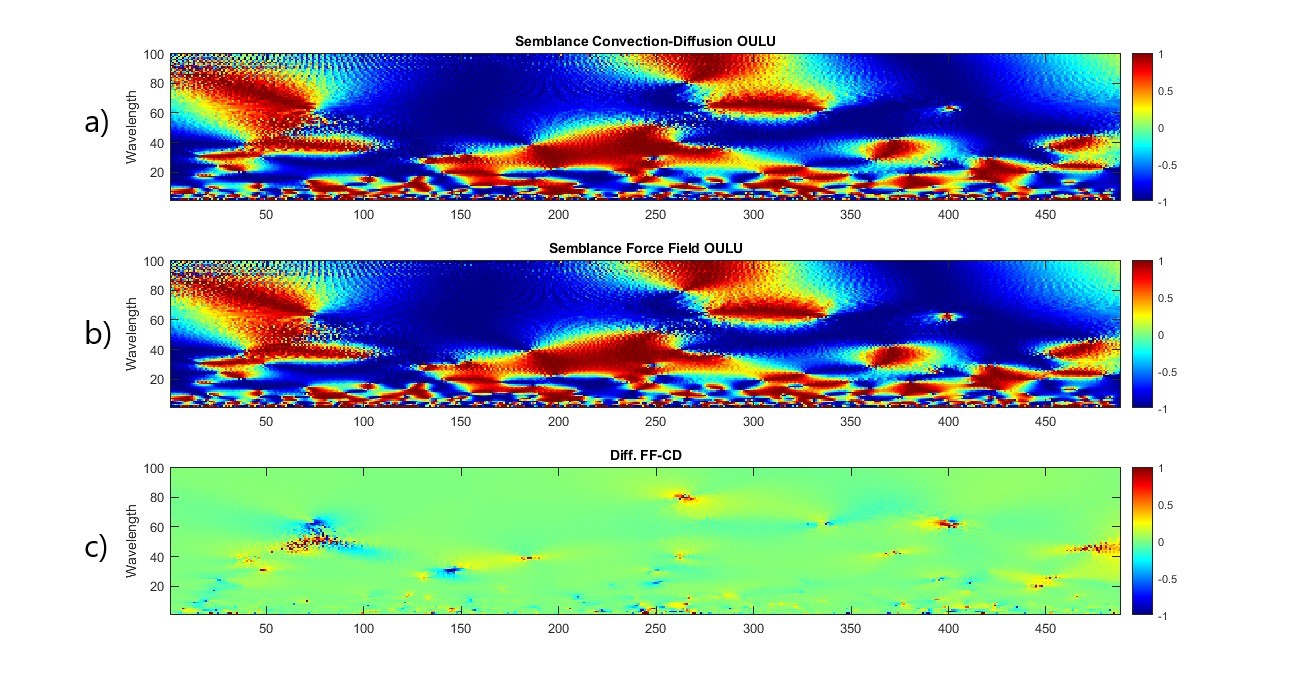}
    \caption{OULU Station a) Semblance between Modulation Factor using Convection-Diffusion with Lagner, Potgieter and Webber LIS in 2003 vs Mean Magnetic Field data. b) Semblance between Modulation Factor using Mean Magnetic Field with Lagner, Potgieter and Webber LIS in 2003 vs Sunspot data. c) Difference between the two previous semblances.}
    \label{sem5}
\end{figure}

 \begin{figure}
     \centering
     \includegraphics[width=0.5\textwidth]{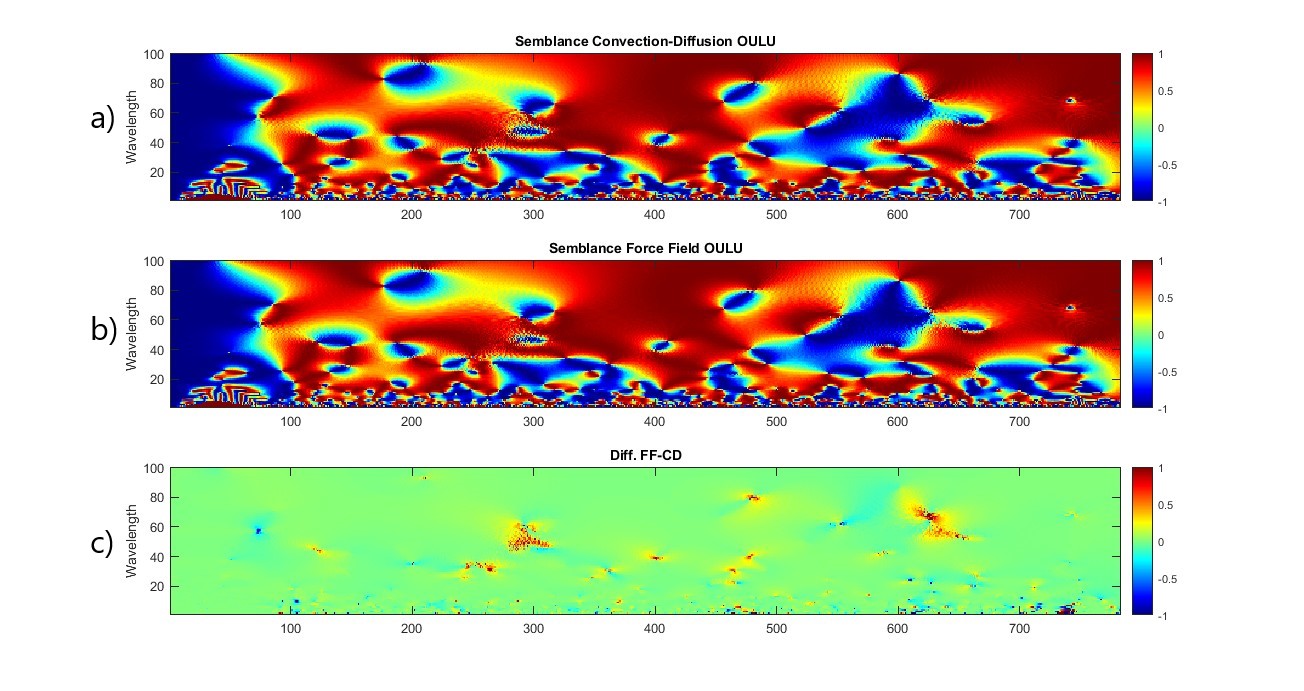}
     \caption{OULU Station a) Semblance between Modulation Factor using Convection-Diffusion with Burguer and Potgieter LIS in 2000 vs Sunspot data. b) Semblance between Modulation Factor using Force Field with Burguer and Potgieter LIS in 2000 vs Sunspot data. c) Difference between the two previous semblances.}
     \label{semOULU17}
 \end{figure}

  \begin{figure}
     \centering
     \includegraphics[width=0.5\textwidth]{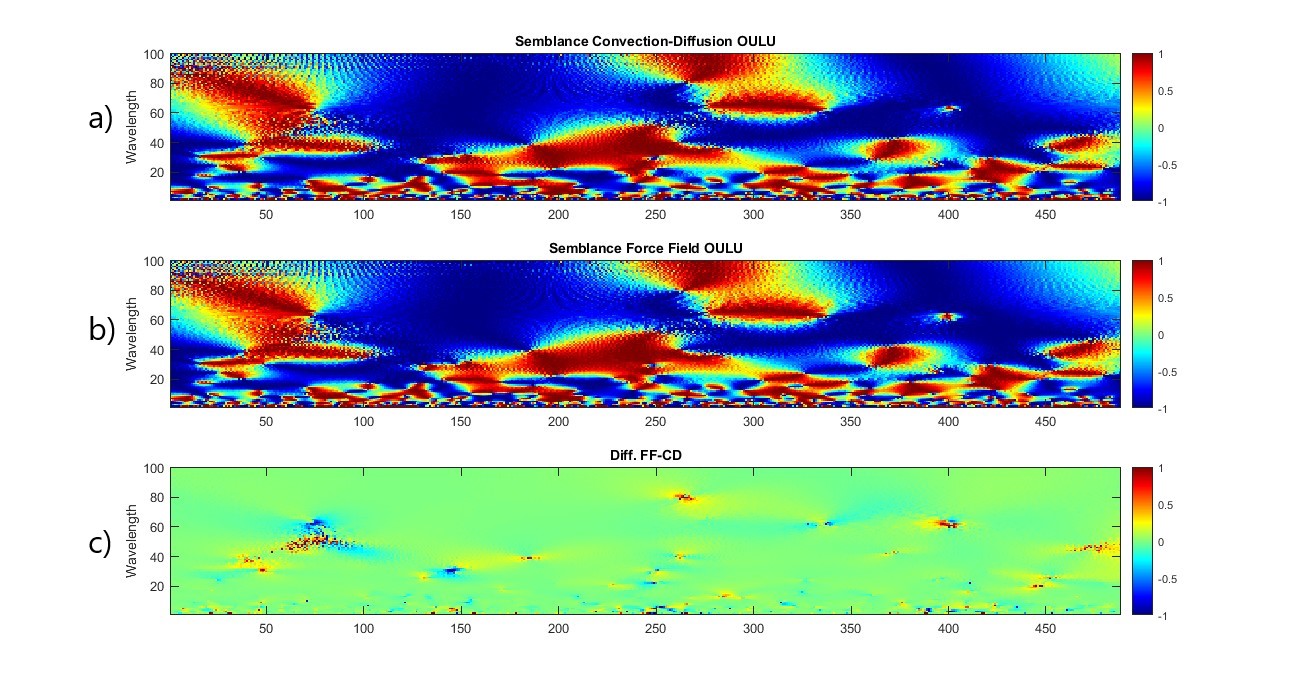}
     \caption{OULU Station a) Semblance between Modulation Factor using Convection-Diffusion with Burguer and Potgieter LIS in 2000 vs Mean Magnetic Field data. b) Semblance between Modulation Factor using Force Field with Burguer and Potgieter LIS in 2000 vs Mean Magnetic Field data. c) Difference between the two previous semblances.}
     \label{sem8}
 \end{figure}

\begin{figure}
    \centering
    \includegraphics[width=0.5\textwidth]{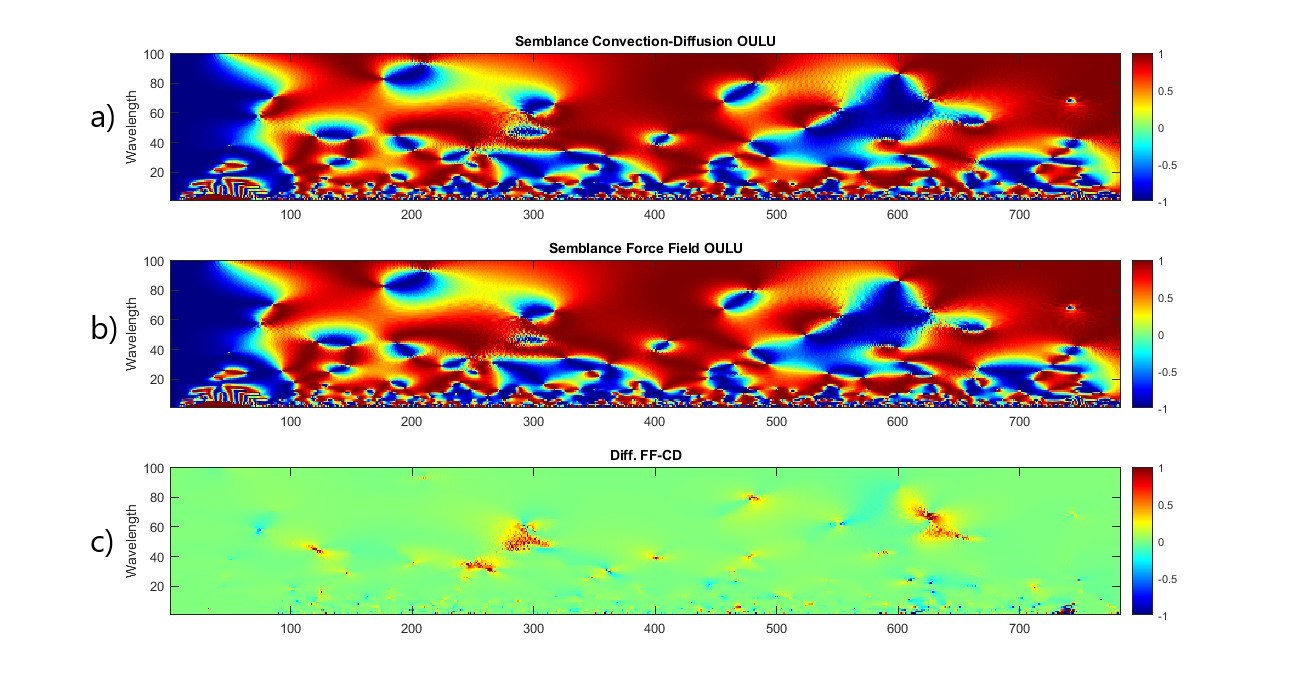}
    \caption{OULU Station a) Semblance between Modulation Factor using Convection-Diffusion with Garcia-Munoz, Mason and Simpson LIS in 1975 vs Sunspot data. b) Semblance between Modulation Factor using Force Field with Garcia-Munoz, Mason and Simpson LIS in 1975 vs Sunspot data. c) Difference between the two previous semblances.}
    \label{sem14}
\end{figure}

\begin{figure}
    \centering
    \includegraphics[width=0.5\textwidth]{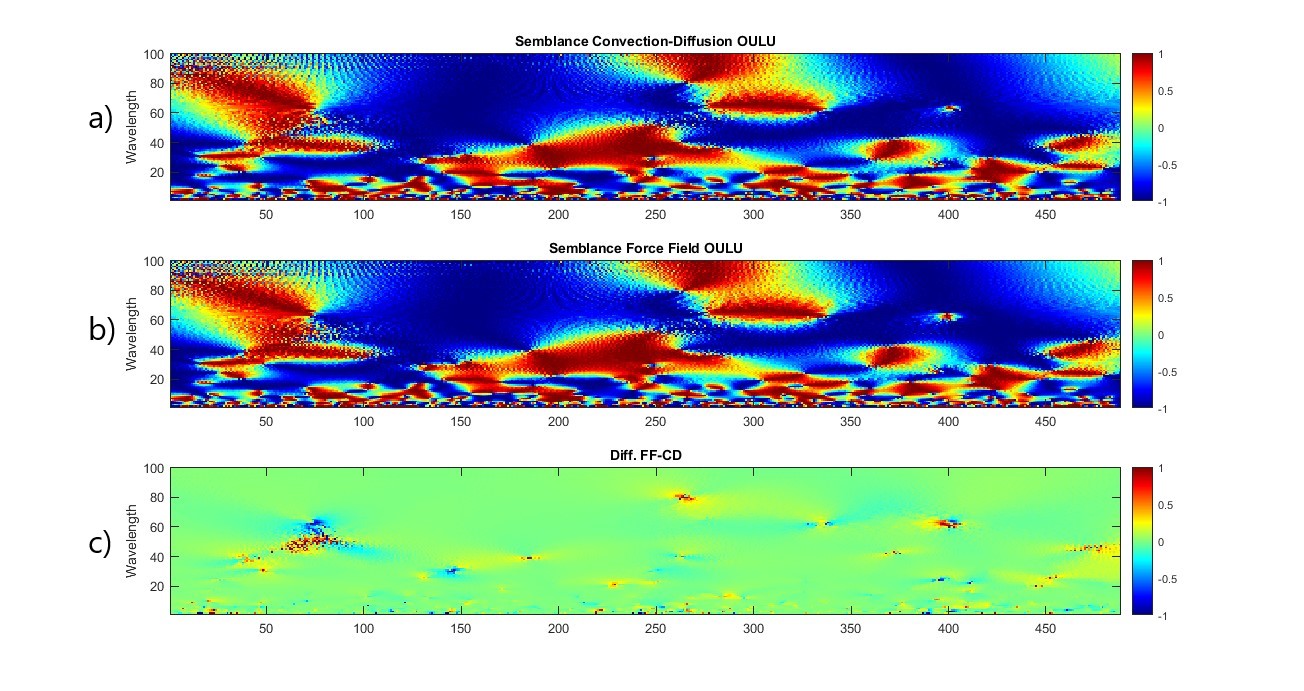}
    \caption{OULU Station a) Semblance between Modulation Factor using Convection-Diffusion with Garcia-Munoz, Mason and Simpson LIS in 1975 vs Mean Magnetic Field data. b) Semblance between Modulation Factor using Force Field with Garcia-Munoz, Mason and Simpson LIS in 1975 vs Mean Magnetic Field data. c) Difference between the two previous semblances.}
    \label{sem17}
\end{figure}

\begin{figure}
    \centering
    \includegraphics[width=0.5\textwidth]{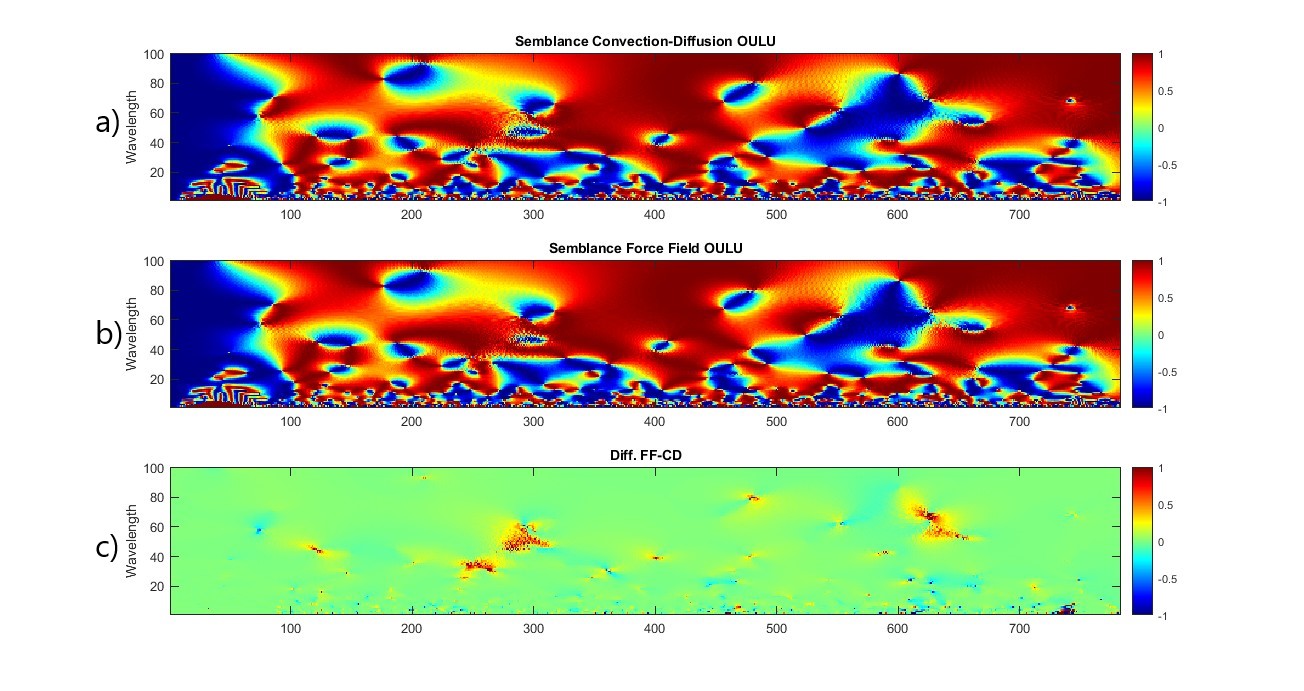}
    \caption{OULU Station a) Semblance between Modulation Factor using Convection-Diffusion with Ghelfi, Barao, Derome and Maurin LIS in 2017 vs Sunspots data. b) Semblance between Modulation Factor using Force Field with Ghelfi, Barao, Derome and Maurin LIS in 2017 vs Sunspots data. c) Difference between the two previous semblances.}
    \label{sem20}
\end{figure}

\begin{figure}
    \centering
    \includegraphics[width=0.5\textwidth]{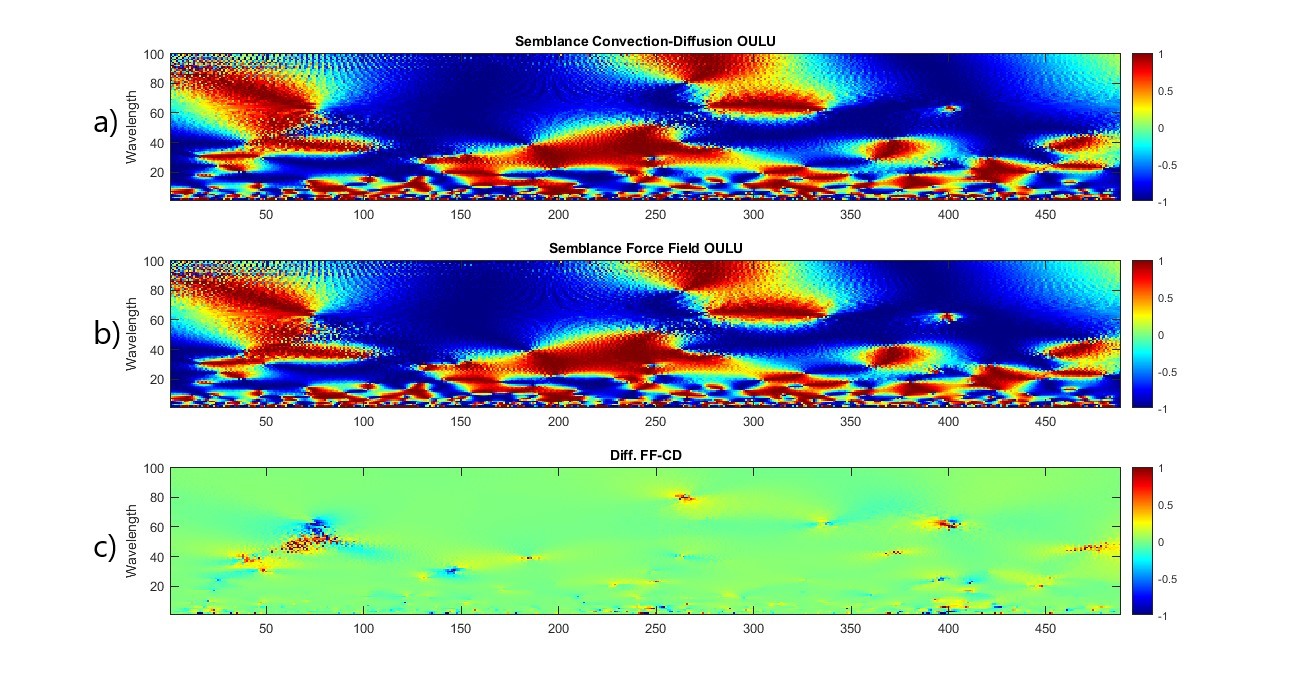}
    \caption{OULU Station a) Semblance between Modulation Factor using Convection-Diffusion with Ghelfi, Barao, Derome and Maurin LIS in 2017 vs Mean Magnetic Field data. b) Semblance between Modulation Factor using Force Field with Ghelfi, Barao, Derome and Maurin LIS in 2017 vs Mean Magnetic Field data. c) Difference between the two previous semblances.}
    \label{sem23}
\end{figure}

The computed models offer similar results among them (figures \(a)\) and \(b)\) for each semblance figure). They even appear as if they were the same model, but thanks to the difference between semblances (figures \(c)\))  it is possible to observe small differences, mainly in Figures \ref{sem17} and \ref{sem14}. If both figures previously mentioned are compared with the rest of the respective semblances in this section, the main difference is in the interval 50 to 100 in number of data and 40 to 80 in the wavelength, where a blue spot changes its size and shape. 

There is correlation between sunspots and modulation phenomena, which can be observed in the figures \ref{sem2}, \ref{semOULU17}, \ref{sem14} and \ref{sem20} highlighted in red.
There is anti-correlation between the mean solar magnetic field and the modulation phenomena observed in blue. Both behaviors are maintained independently of the magnetospheric latitude.

\hypertarget{second-analysis}{%
\section{\texorpdfstring{\textbf{Second
Analysis}}{Second Analysis}}\label{Second-analysis}}
The same calculations were performed, but the modulation factors were averaged. The data used for the following analysis were: KERG, MOSC, OULU, THUL, and HRMS (Table \ref{SA}). This analysis was conducted to generalize the previous results and to analyze which LIS have the best fit.

\begin{table}[!ht]
\begin{tabular}{|c|c|}
\hline
\textbf{Neutron Monitor} & \textbf{Cut-off Rigidity {[}GeV{]}} \\ \hline
\textbf{Greenland (THUL)}            & \textbf{0.3}                               \\ \hline
\textbf{Finland (OULU)}            & \textbf{0.81}                              \\ \hline
\textbf{Kerguelen (KERG)}            & \textbf{1.14}                              \\ \hline
\textbf{Moscow (MOSC)}            & \textbf{2.43}                              \\ \hline
\textbf{Hermanus (HRMS)}            & \textbf{4.58}                              \\ \hline
\end{tabular}
\caption{Cut-off Rigidity from the used stations in the second analysis}
\label{SA}
\end{table}

The Figure \ref{StationsWorld} are the position of the used stations in both analyses.
\begin{figure}[!ht]
    \centering
    \includegraphics[width=0.5 \textwidth]{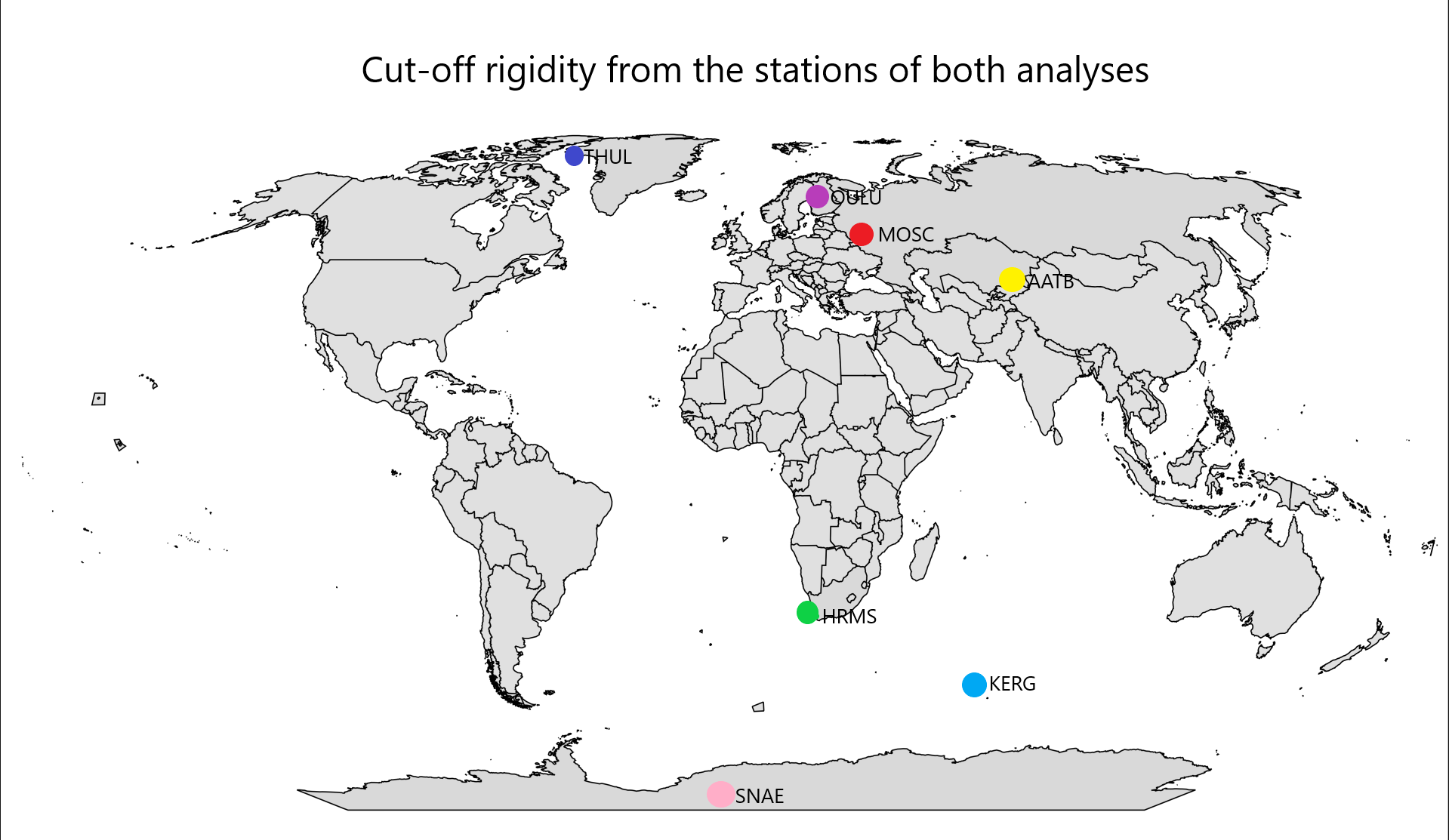}
    \caption{Position of the neutron monitors around the world used in both analyses}
    \label{StationsWorld}
\end{figure}

\begin{figure}[ht!]
    \centering    
    \includegraphics[width=0.37\textwidth]{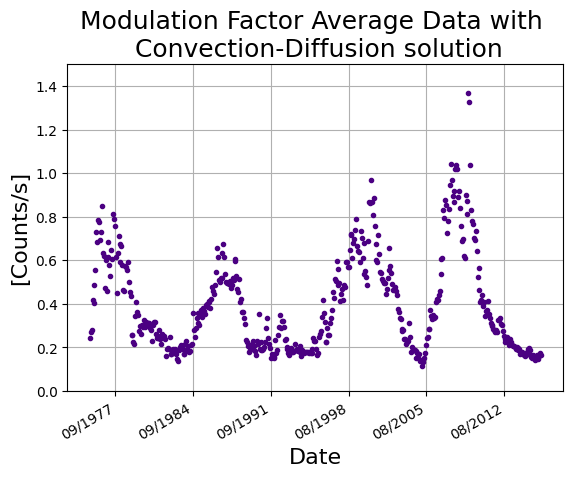}
    \caption{Averaged Modulation Factor among KERG, HRMS, MOSC, OULU and THUL stations. Modulation factor with Convection-Diffusion solution using Lagner, Potgieter \& Webber LIS in 2003.}
    \label{fig14}
\end{figure}

\begin{figure}[ht!]
    \centering   
    \includegraphics[width=0.37\textwidth]{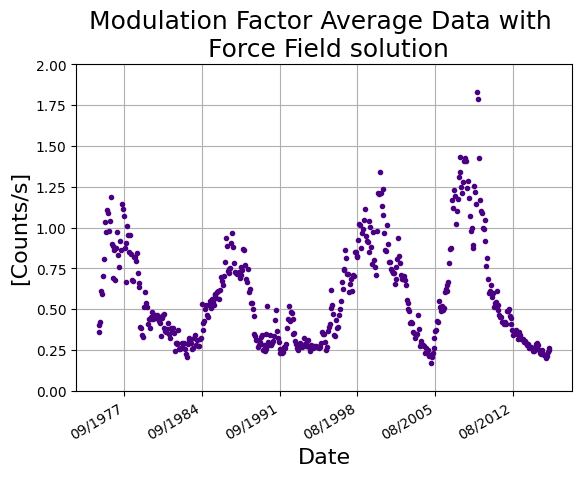}
    \caption{Averaged Modulation Factor among KERG, HRMS, MOSC, OULU and THUL stations. Modulation factor with Force Field solution using Lagner, Potgieter \& Webber LIS in 2003..}
    \label{fig15}
\end{figure}

\begin{figure}[ht!]
    \centering    
    \includegraphics[width=0.37\textwidth]{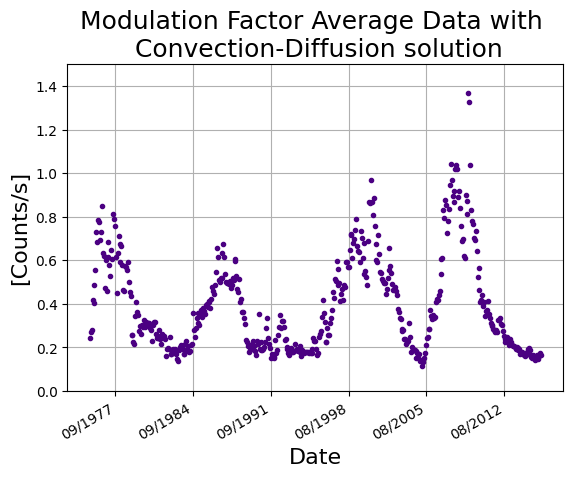}
    \caption{Averaged Modulation Factor among KERG, HRMS, MOSC, OULU and THUL stations. Modulation factor with Convection-Diffusion solution using Burguer and Potgieter LIS in 2000.}
    \label{fig16}
\end{figure}

\begin{figure}[ht!]
    \centering  
    \includegraphics[width=0.37\textwidth]{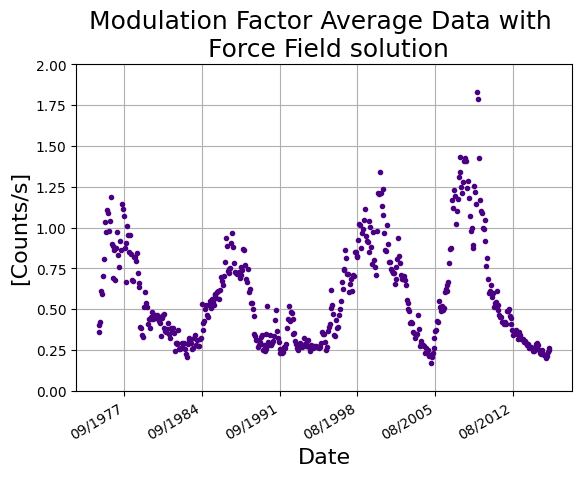}
    \caption{Averaged Modulation Factor among KERG, HRMS, MOSC, OULU and THUL stations. Modulation factor with Force Field solution using Burguer and Potgieter LIS in 2000.}
    \label{fig17}
\end{figure}

\begin{figure}[ht!]
    \centering   
    \includegraphics[width=0.37\textwidth]{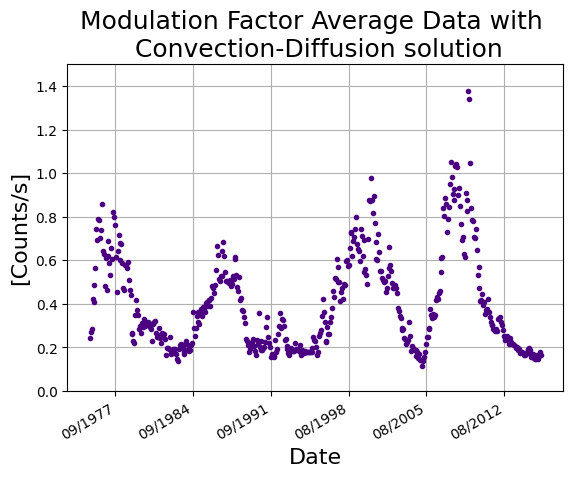}
    \caption{Averaged Modulation Factor among KERG, HRMS, MOSC, OULU and THUL stations. Modulation factor with Convection-Diffusion solution using Garcia-Munoz, Mason and Simpson LIS in 1975.}
    \label{fig18}
\end{figure}

\begin{figure}[ht!]
    \centering 
    \includegraphics[width=0.37\textwidth]{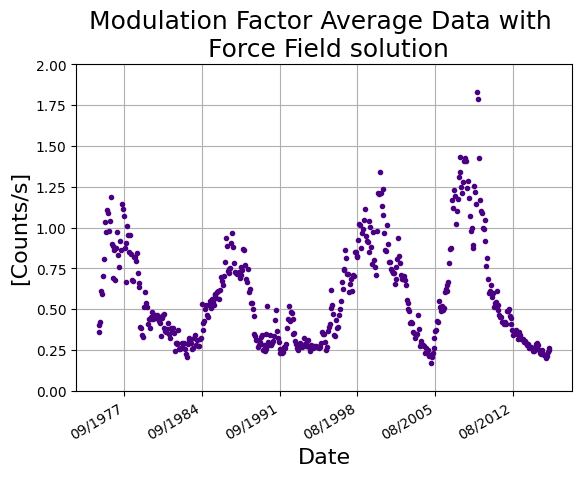}
    \caption{Averaged Modulation Factor among KERG, HRMS, MOSC, OULU and THUL stations. Modulation factor with Force Field solution using Garcia-Munoz, Mason and Simpson LIS in 1975.}
    \label{fig19}
\end{figure}

\begin{figure}[ht!]
    \centering  
    \includegraphics[width=0.37\textwidth]{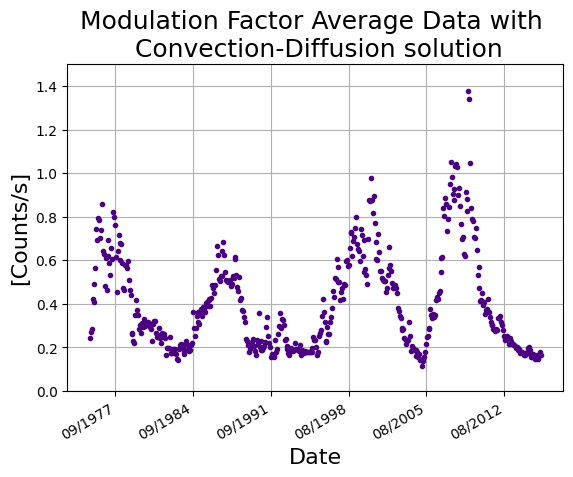}
    \caption{Averaged Modulation Factor among KERG, HRMS, MOSC, OULU and THUL stations. Modulation factor with Convection-Diffusion solution using Ghelfi, Barao, Derome and Maurin LIS in 2017.}
    \label{fig20}
\end{figure}

\begin{figure}[ht!]
    \centering  
    \includegraphics[width=0.37\textwidth]{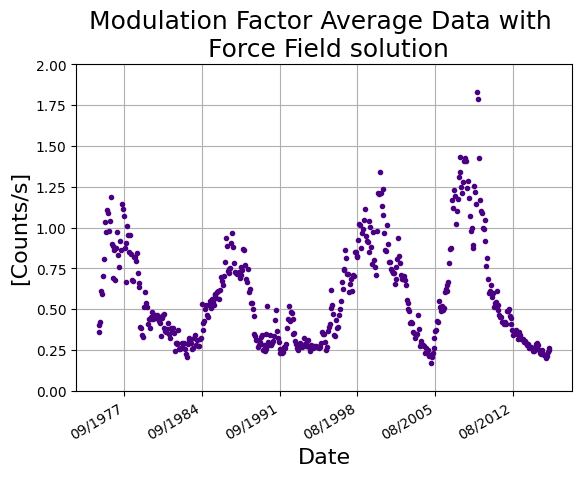}
    \caption{Averaged Modulation Factor among KERG, HRMS, MOSC, OULU and THUL stations. Modulation factor with Force Field solution using Ghelfi, Barao, Derome and Maurin LIS in 2017.}
    \label{fig21}
\end{figure}

The Figures \ref{fig14} to \ref{fig21} are the modulation factors averaged  using convection-diffusion and force field solutions with Lagner, Potgieter \& Webber LIS in 2003, Burguer \& Potgieter LIS in 2000, Garcia-Munoz, Mason \& Simpson LIS in 1975 and Ghelfi, Barao, Derome \& Maurin LIS in 2017.

\begin{figure}
    \centering
    \includegraphics[width=0.5\textwidth]{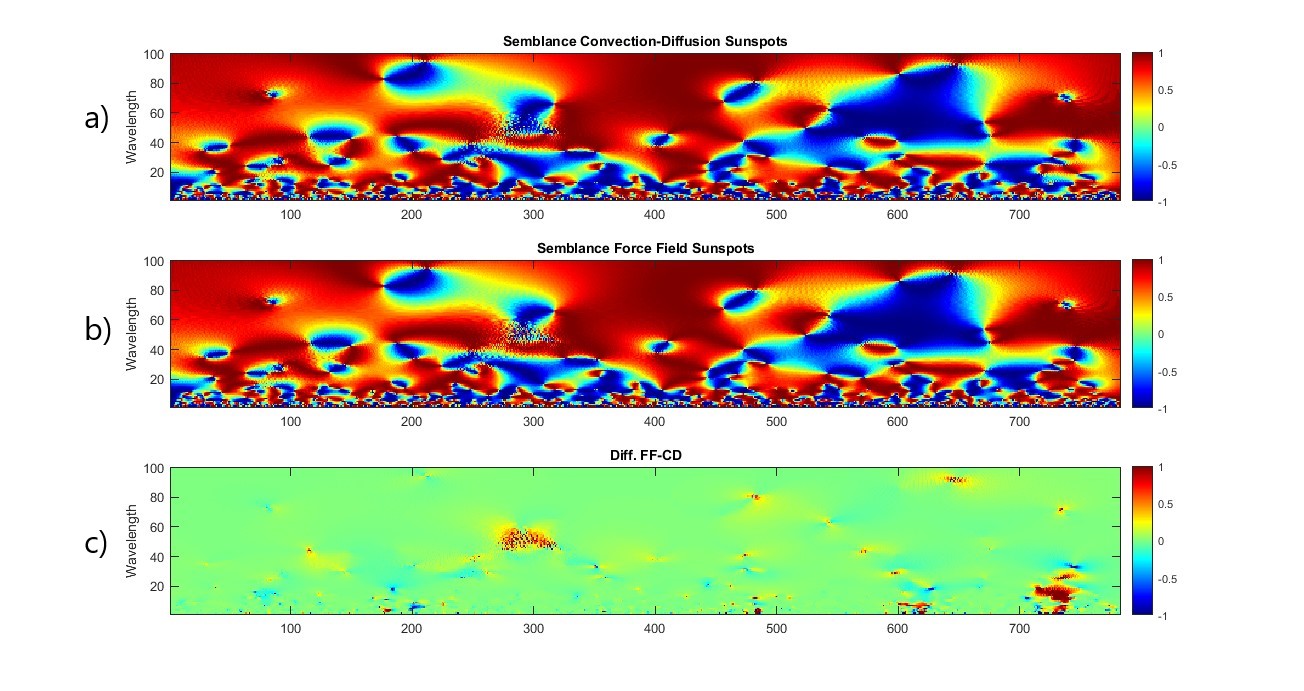}
    \caption{Average among HRMS, KERG, MOSC, OULU and THUL a) Semblance between Average Modulation Factor using Convection-Diffusion with Lagner, Potgieter and Webber LIS in 2003 vs Sunspots data. b) Semblance between Average Modulation Factor using Force Field with Lagner, Potgieter and Webber LIS in 2003 vs Sunstpots data. c) Difference between the two previous semblances.}
    \label{sem25}
\end{figure}

\begin{figure}
    \centering
    \includegraphics[width=0.5\textwidth]{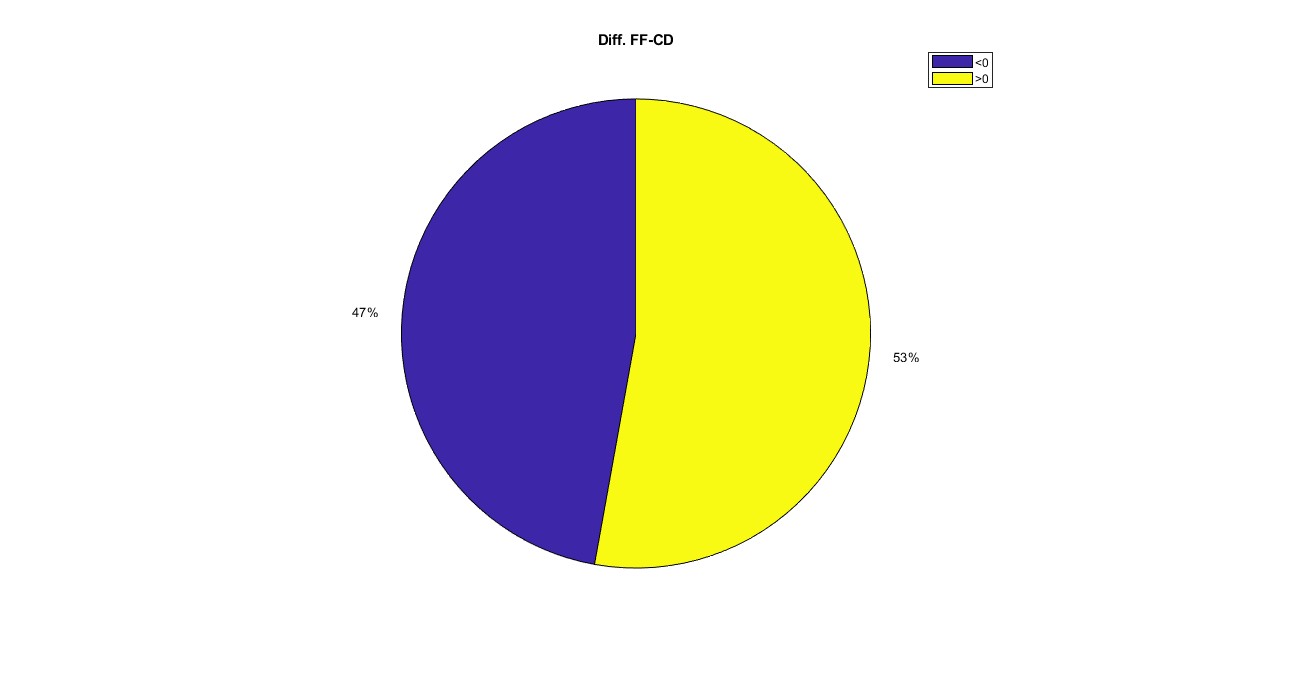}
    \caption{Contained zeros into Figure \ref{sem25} c).}
    \label{sem26}
\end{figure}

\begin{figure}
    \centering
    \includegraphics[width=0.5\textwidth]{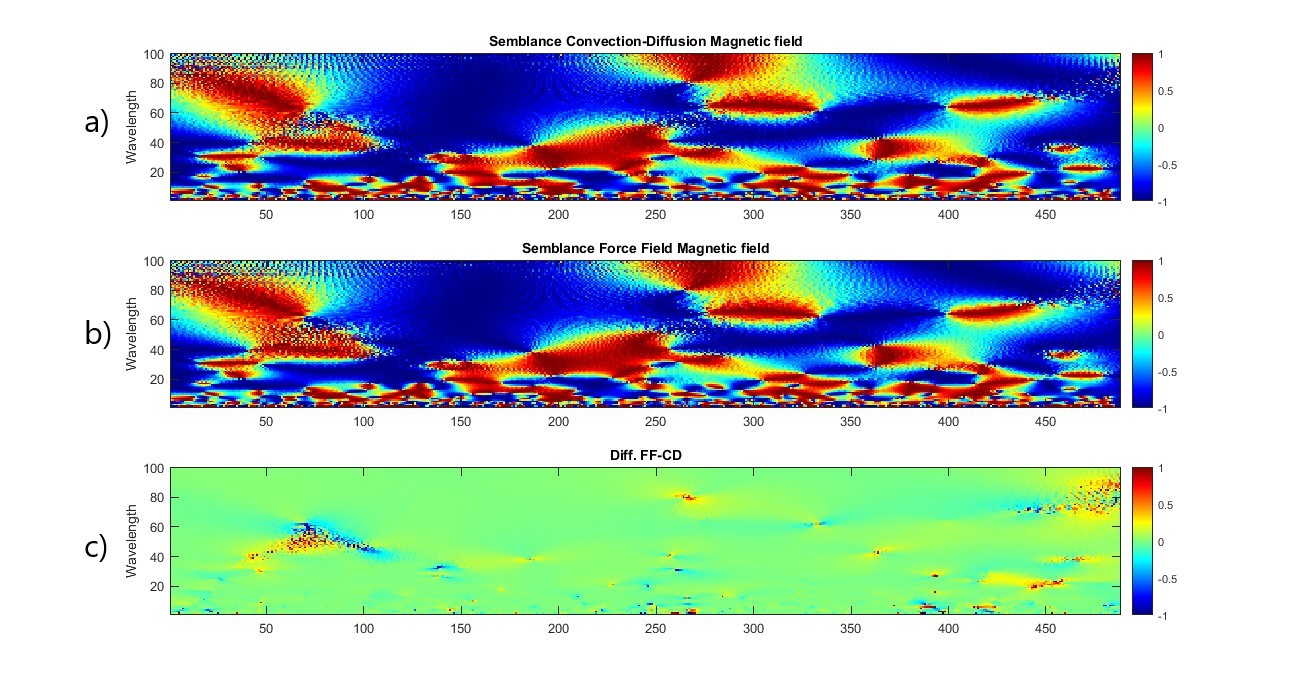}
    \caption{Average among HRMS, KERG, MOSC, OULU and THUL a) Semblance between Average Modulation Factor using Convection-Diffusion with Lagner, Potgieter and Webber LIS in 2003 vs Solar Magnetic Field data. b) Semblance between Average Modulation Factor using Force Field with Lagner, Potgieter and Webber LIS in 2003 vs Solar Magnetic Field data. c) Difference between the two previous semblances.}
    \label{sem27}
\end{figure}

\begin{figure}
    \centering
    \includegraphics[width=0.5\textwidth]{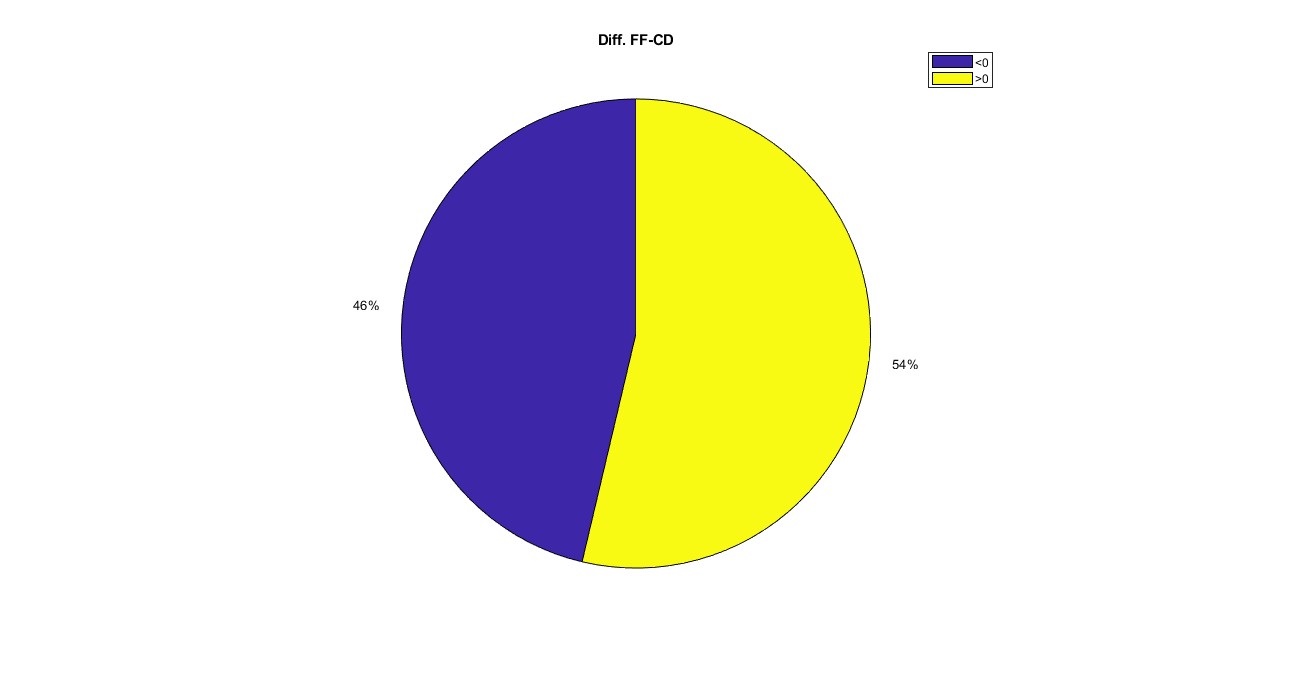}
    \caption{Contained zeros into Figure \ref{sem27} c).}
    \label{sem28}
\end{figure}

\begin{figure}
    \centering
    \includegraphics[width=0.5\textwidth]{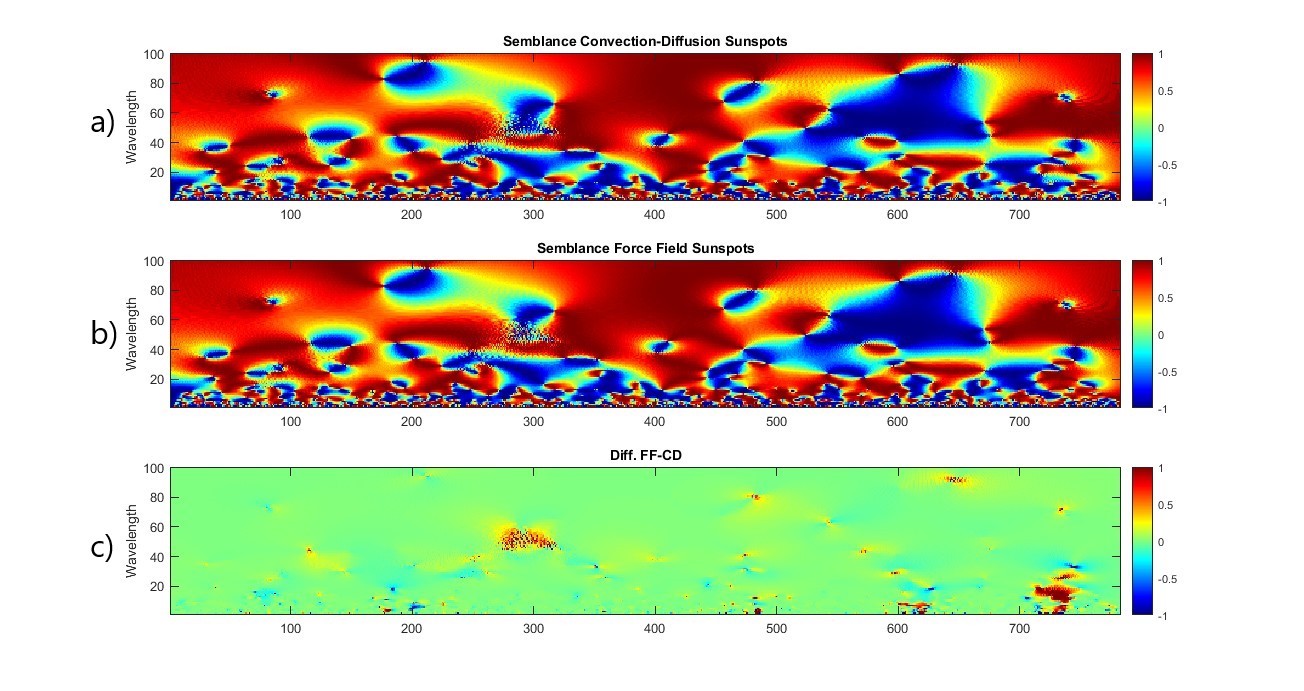}
    \caption{Average among HRMS, KERG, MOSC, OULU and THUL a) Semblance between Average Modulation Factor using Convection-Diffusion with Burguer and Potgieter LIS in 2000 vs Sunspots data. b) Semblance between Average Modulation Factor using Force Field with Burguer and Potgieter LIS in 2000 vs Sunstpots data. c) Difference between the two previous semblances.}
    \label{sem29}
\end{figure}

\begin{figure}
    \centering
    \includegraphics[width=0.5\textwidth]{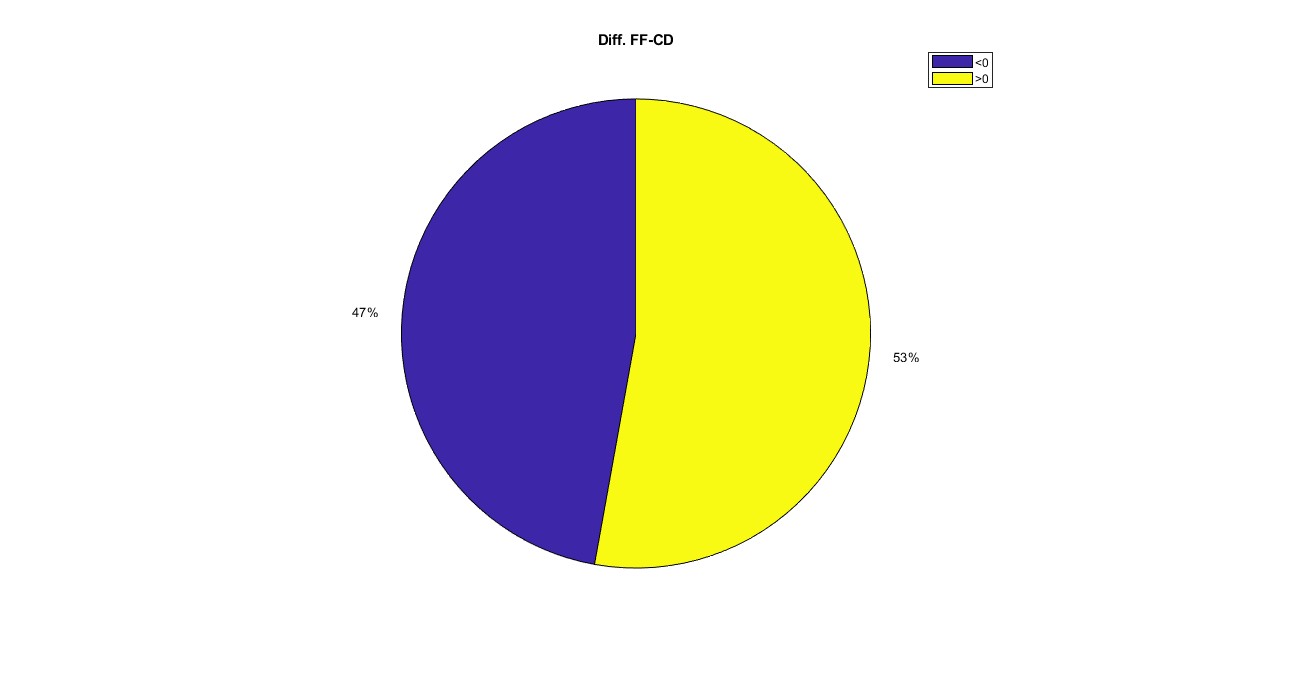}
    \caption{Contained zeros into Figure \ref{sem29} c).}
    \label{sem30}
\end{figure}

\begin{figure}
    \centering
    \includegraphics[width=0.5\textwidth]{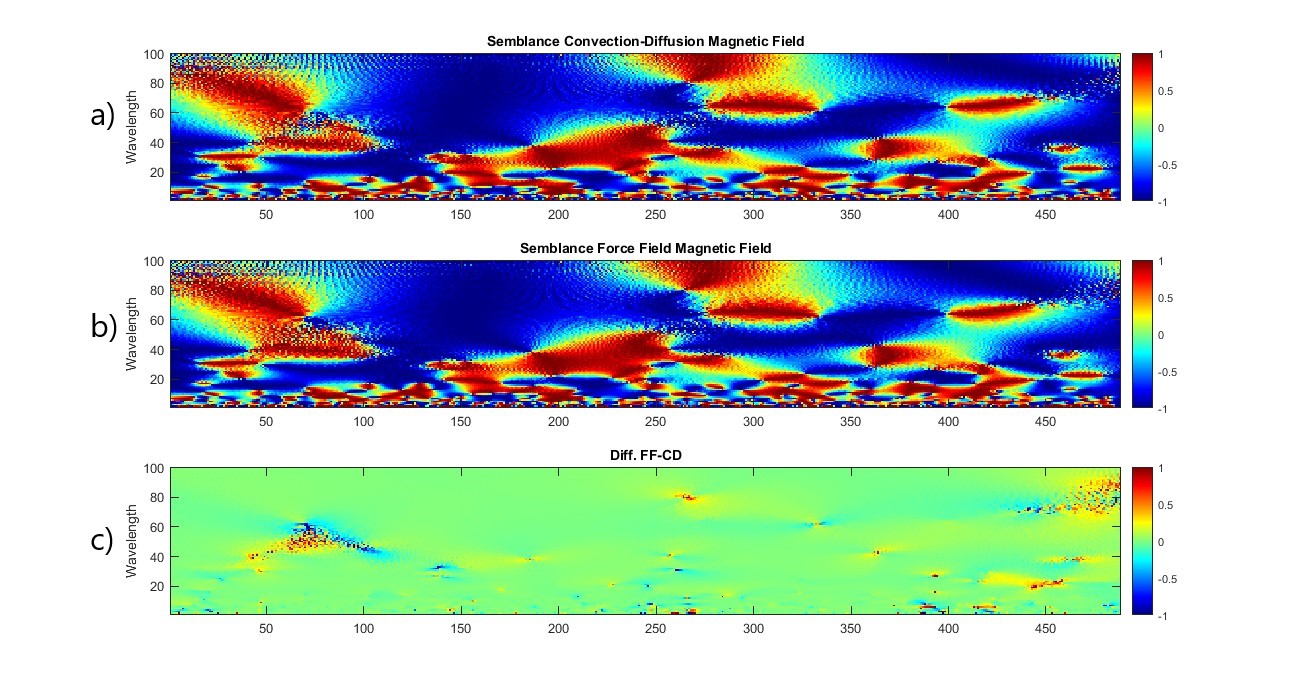}
    \caption{Average among HRMS, KERG, MOSC, OULU and THUL a) Semblance between Average Modulation Factor using Convection-Diffusion with Burguer and Potgieter LIS in 2000 vs Solar Magnetic Field data. b) Semblance between Average Modulation Factor using Force Field with Burguer and Potgieter LIS in 2000 vs Solar Magnetic Field data. c) Difference between the two previous semblances.}
    \label{sem31}
\end{figure}

\begin{figure}
    \centering
    \includegraphics[width=0.5\textwidth]{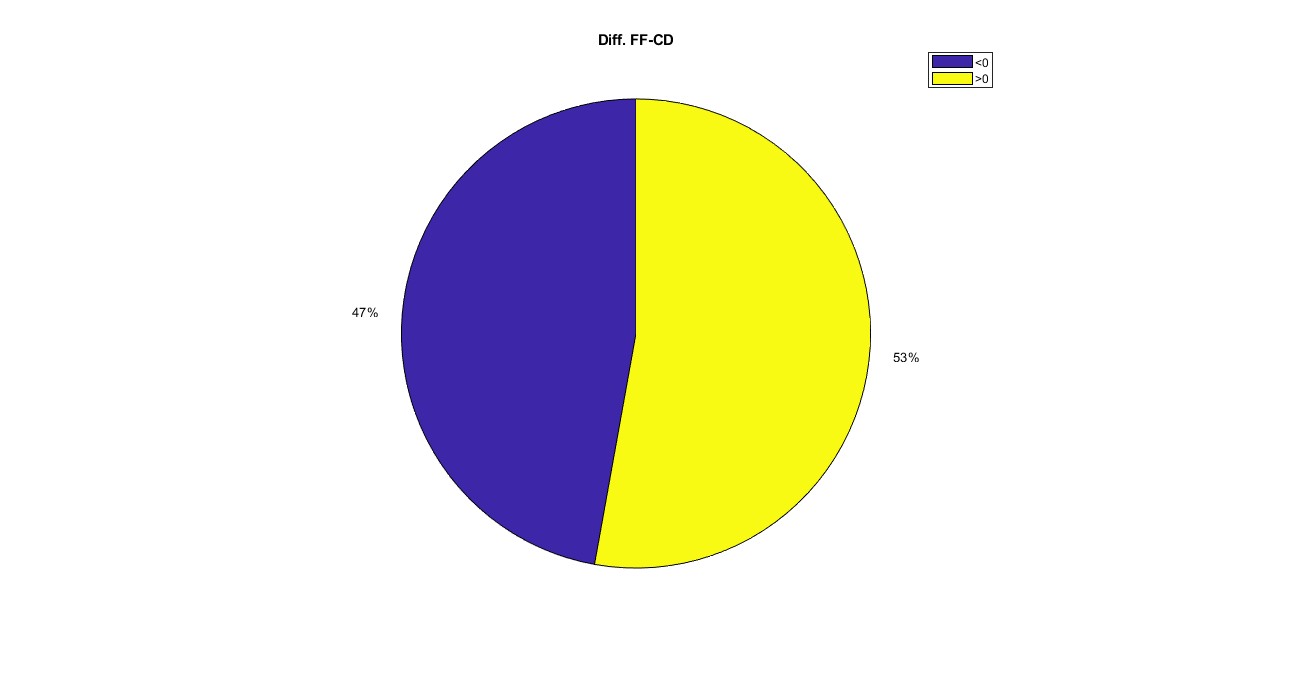}
    \caption{Contained zeros into Figure \ref{sem31} c).}
    \label{sem32}
\end{figure}

\begin{figure}
    \centering
    \includegraphics[width=0.5\textwidth]{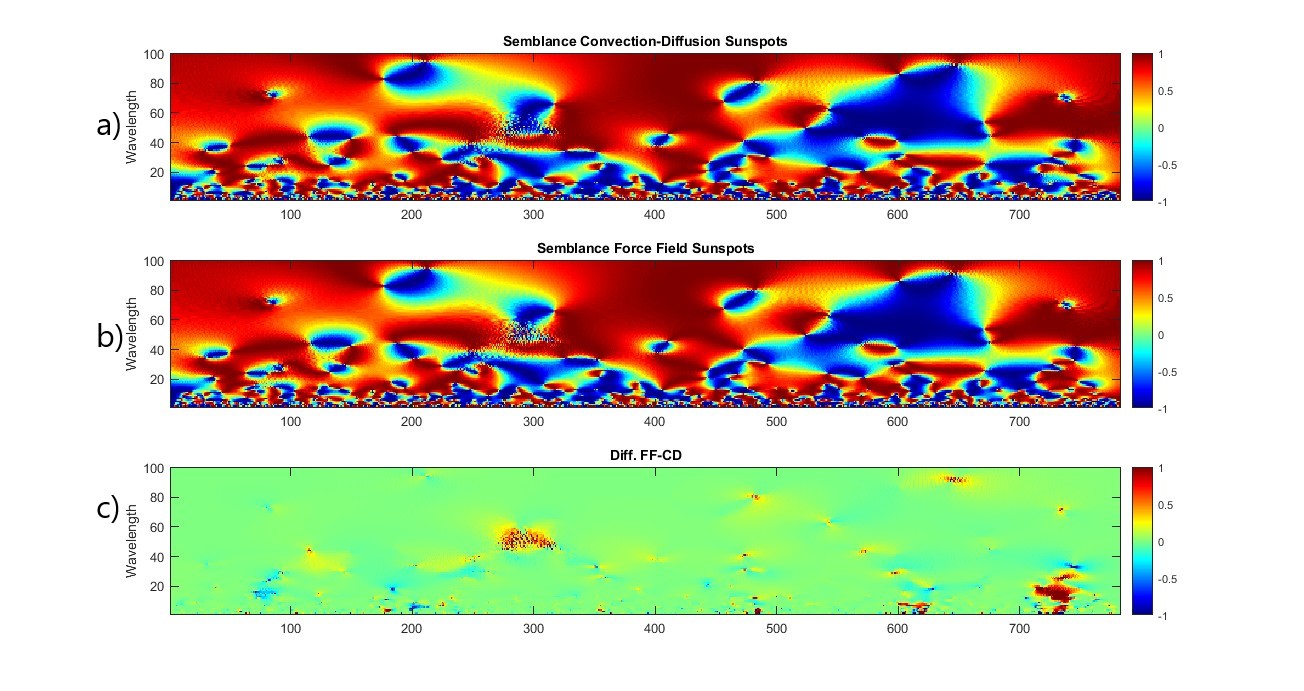}
    \caption{Average among HRMS, KERG, MOSC, OULU and THUL a) Semblance between Average Modulation Factor using Convection-Diffusion with Garcia-Munoz, Mason and Simpson LIS in 1975 vs Sunspots data. b) Semblance between Average Modulation Factor using Force Field with Garcia-Munoz, Mason and Simpson LIS in 1975 vs Sunstpots data. c) Difference between the two previous semblances.}
    \label{sem33}
\end{figure}

\begin{figure}
    \centering
    \includegraphics[width=0.5\textwidth]{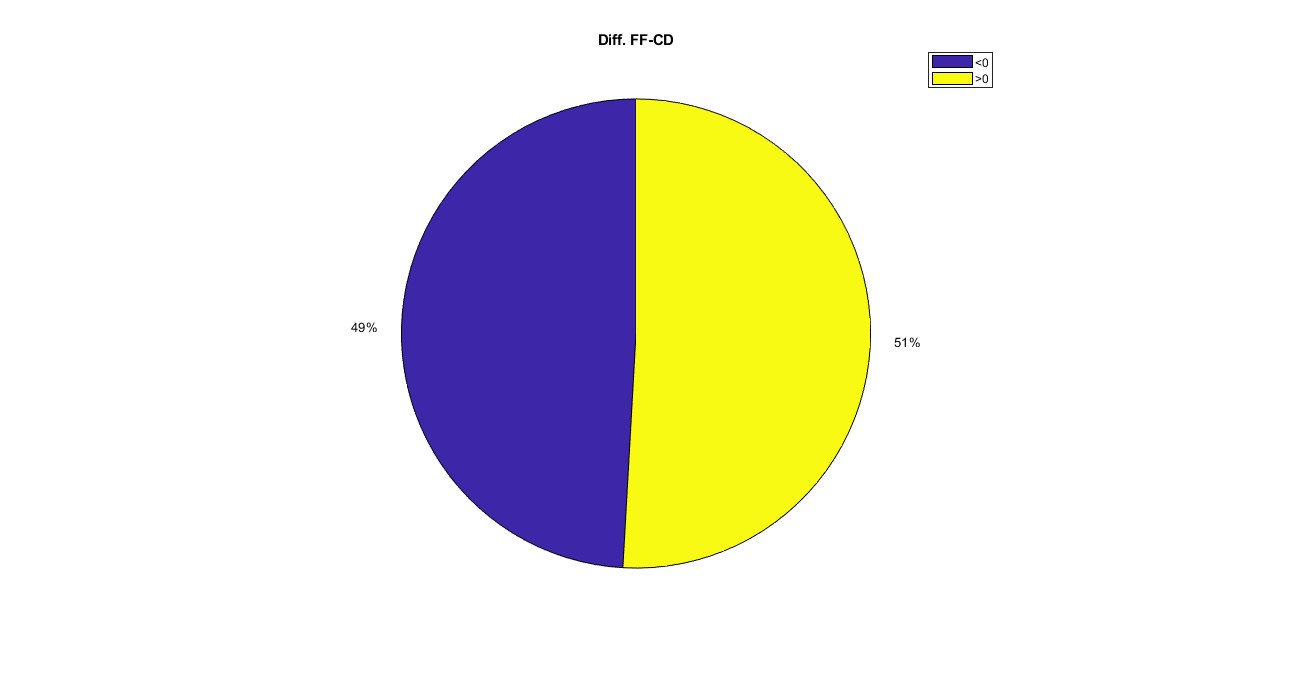}
    \caption{Contained zeros into Figure \ref{sem33} c).}
    \label{sem34}
\end{figure}

\begin{figure}
    \centering
    \includegraphics[width=0.5\textwidth]{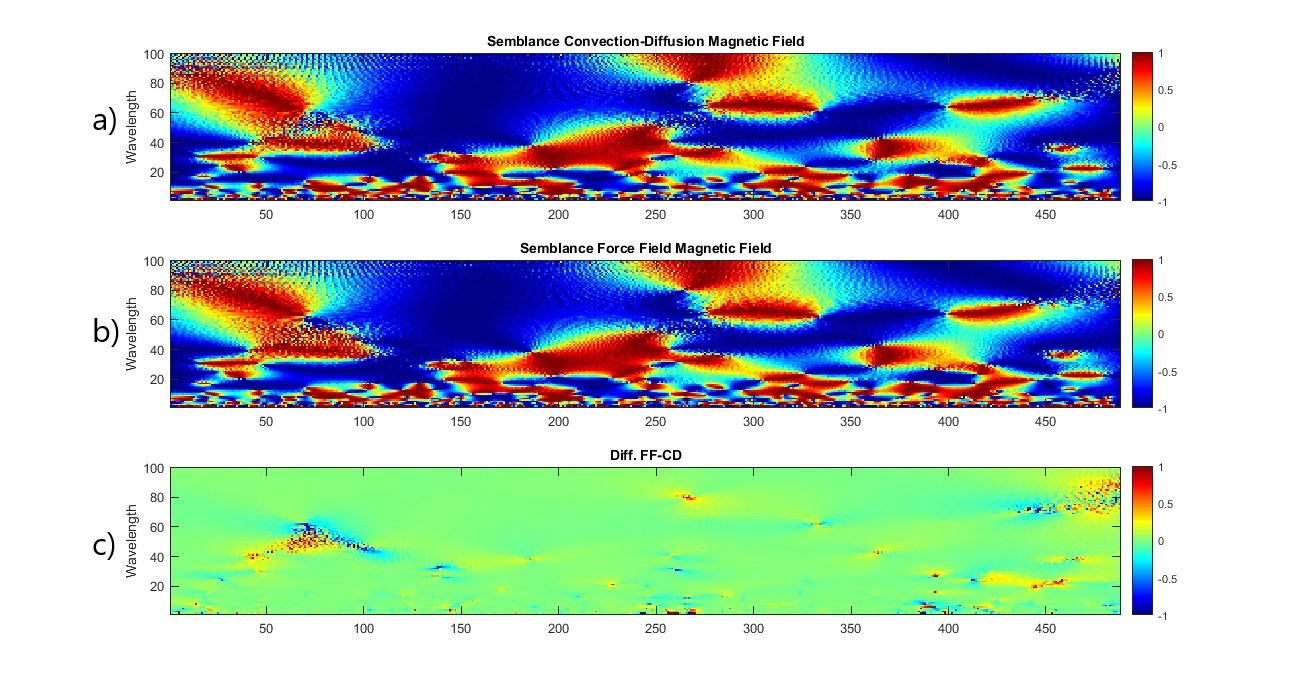}
    \caption{Average among HRMS, KERG, MOSC, OULU and THUL a) Semblance between Average Modulation Factor using Convection-Diffusion with Garcia-Munoz, Mason and Simpson LIS in 1975 vs Solar Magnetic Field data. b) Semblance between Average Modulation Factor using Force Field with Garcia-Munoz, Mason and Simpson LIS in 1975 vs Solar Magnetic Field data. c) Difference between the two previous semblances.}
    \label{sem35}
\end{figure}

\begin{figure}
    \centering
    \includegraphics[width=0.5\textwidth]{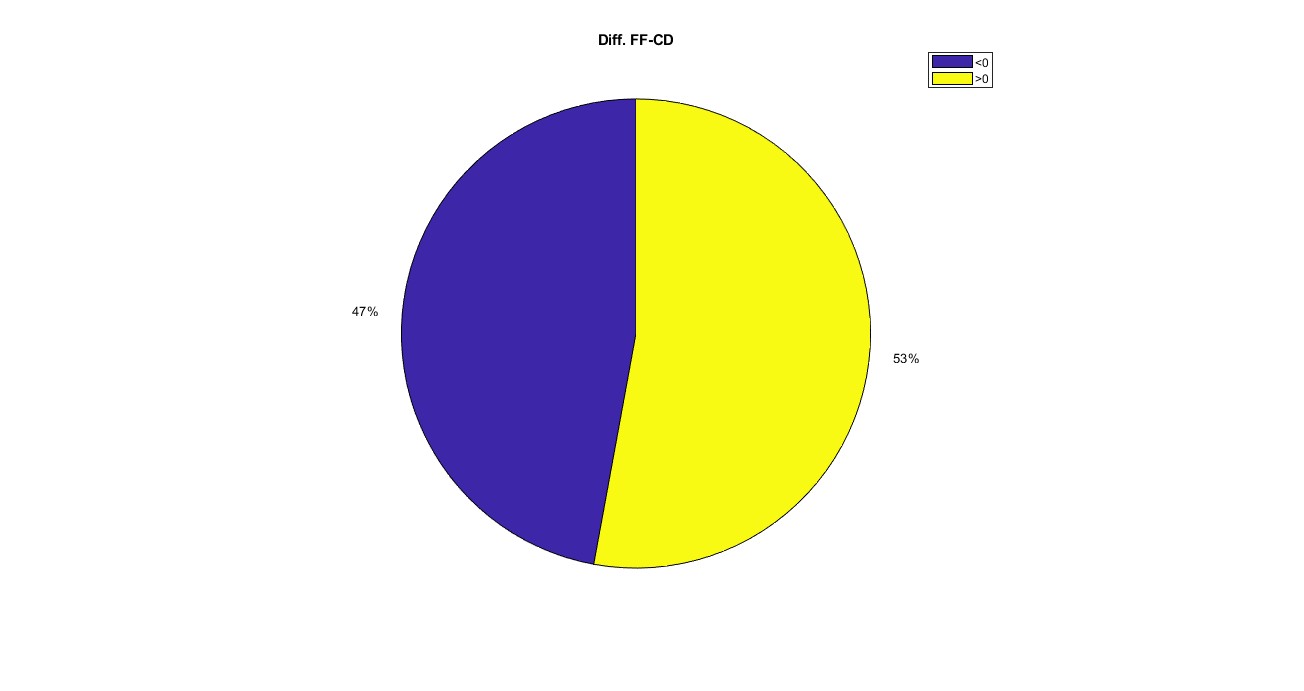}
    \caption{Contained zeros into Figure \ref{sem35} c).}
    \label{sem36}
\end{figure}

\begin{figure}
    \centering
    \includegraphics[width=0.5\textwidth]{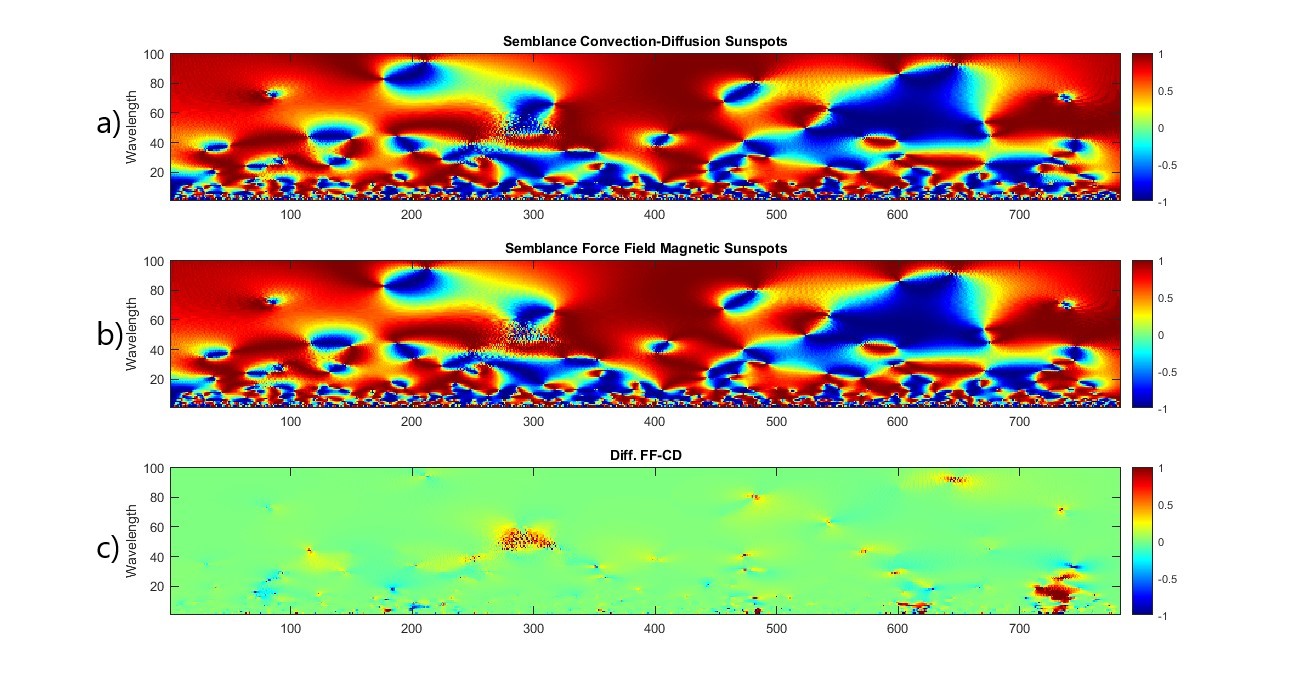}
    \caption{Average among HRMS, KERG, MOSC, OULU and THUL a) Semblance between Average Modulation Factor using Convection-Diffusion with Ghelfi, Barao, Derome and Maurin LIS in 2017 vs Sunspots data. b) Semblance between Average Modulation Factor using Force Field with Ghelfi, Barao, Derome and Maurin LIS in 2017 vs Sunstpots data. c) Difference between the two previous semblances.}
    \label{sem37}
\end{figure}

\begin{figure}
    \centering
    \includegraphics[width=0.5\textwidth]{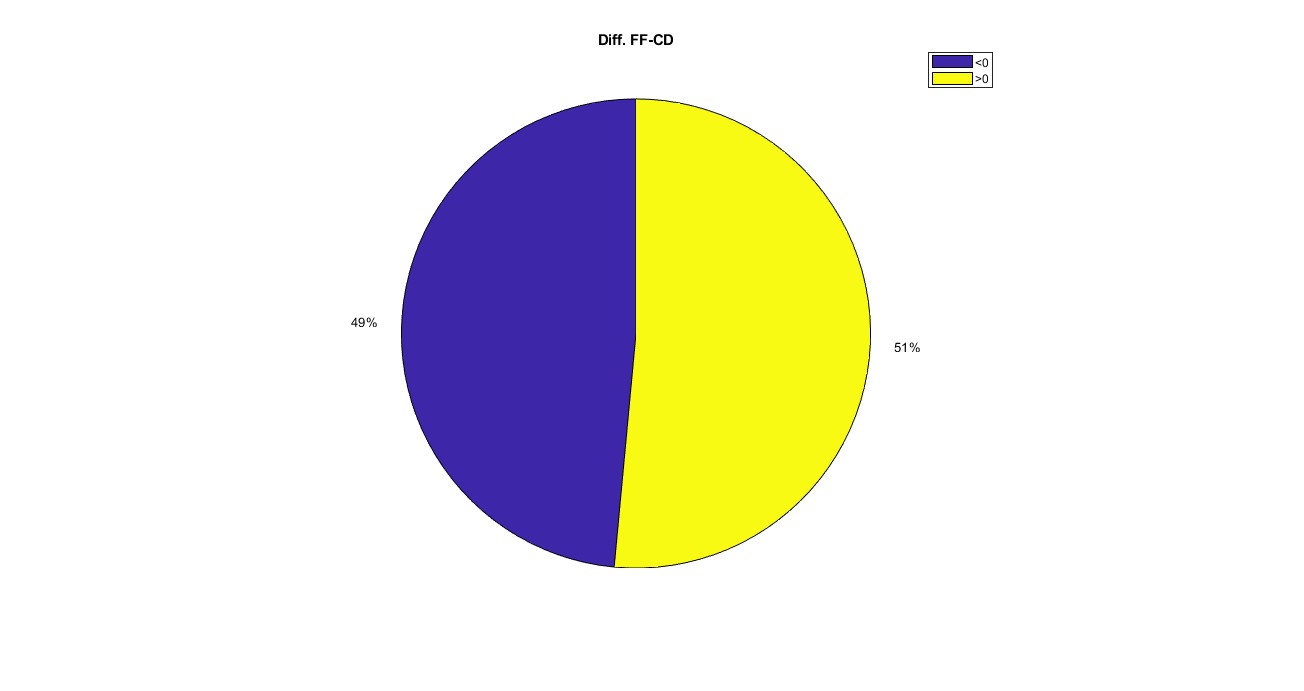}
    \caption{Contained zeros into Figure \ref{sem37} c).}
    \label{sem38}
\end{figure}

\begin{figure}
    \centering
    \includegraphics[width=0.5\textwidth]{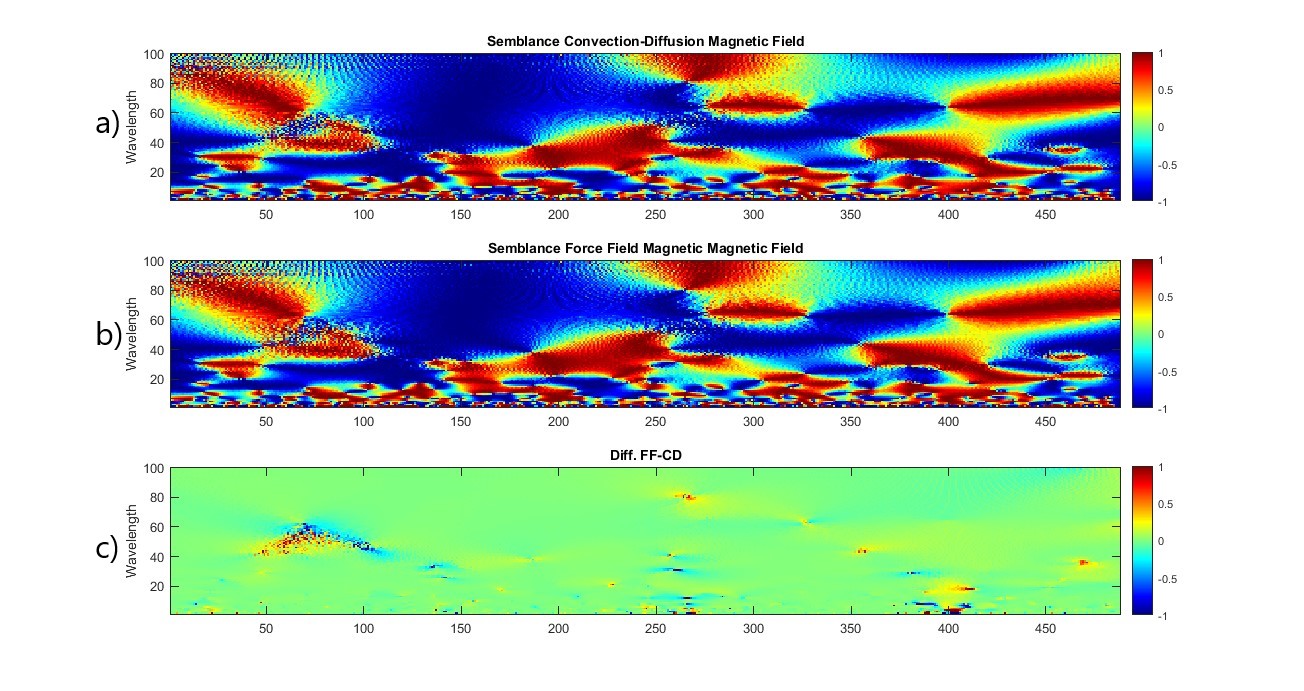}
    \caption{Average among HRMS, KERG, MOSC, OULU and THUL a) Semblance between Average Modulation Factor using Convection-Diffusion with Ghelfi, Barao, Derome and Maurin LIS in 2017 vs Solar Magnetic Field data. b) Semblance between Average Modulation Factor using Force Field with Ghelfi, Barao, Derome and Maurin LIS in 2017 vs Solar Magnetic Field data. c) Difference between the two previous semblances.}
    \label{sem39}
\end{figure}

\begin{figure}
    \centering
    \includegraphics[width=0.5\textwidth]{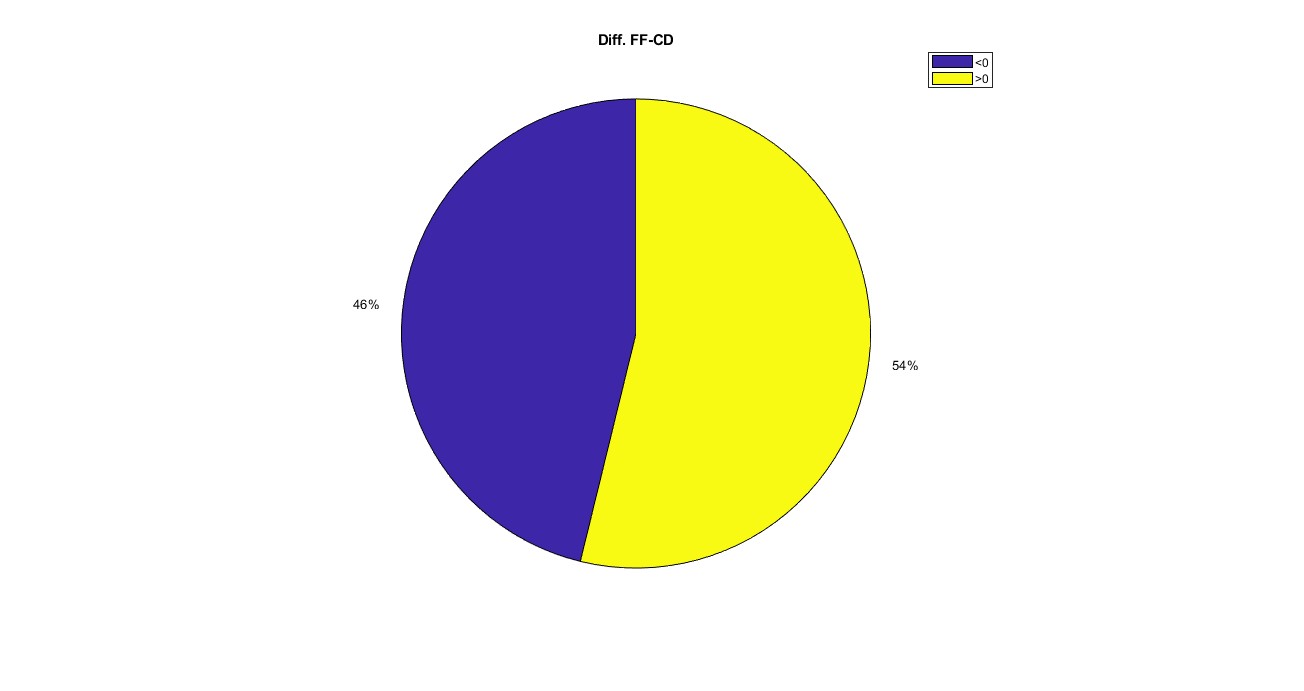}
    \caption{Contained zeros into Figure \ref{sem39} c).}
    \label{sem40}
\end{figure}

The Figures \ref{sem25} to \ref{sem40} show the semblance between the aforementioned modulation factors with sunspots and mean magnetic field data. The figures \ref{sem26}, \ref{sem28}, \ref{sem30}, \ref{sem32} and \ref{sem34} show the percentage of zeros contained in the difference between semblance (Figures \(c)\)). Thanks to this analysis, the LIS models were selected because they offer the most reliable results.

In this case, an average modulation factor of several neutron detectors around the world was computed to analyze  the general analyze  between sunspots and mean solar magnetic field and modulation phenomena.
The results were utterly similar to the first analysis, in fact the analysis of the contained zeros confirms the hypothesis, that there is a correlation between sunspots and modulation phenomena and an anti-correlation between the mean magnetic field and modulation phenomena.
The analysis of the zeros is summarized in Table \ref{analysisZeros}.
\begin{table*}[ht!]
\begin{tabular}{|c|cc|cc|}
\hline
{\textbf{Analysis of contained zeros}} &
  \multicolumn{2}{c|}{\textbf{Sunspots}} &
  \multicolumn{2}{c|}{\textbf{Mean Magnetic Field}} \\ \cline{2-5} 
 &
  \multicolumn{1}{c|}{\textbf{\ \textless{}0} \ } &
  \textbf{\ \textgreater{}0 \ } &
  \multicolumn{1}{c|}{\textbf{\ \ \ \ \ \textless{}0} \ \ \ \ \ } &
  \textbf{\ \textgreater{} 0 \ } \\ \hline
\textbf{Webber \& Lockwood} & \multicolumn{1}{c|}{47} & 53 & \multicolumn{1}{c|}{46} & 54 \\ \hline
\textbf{Burguer}            & \multicolumn{1}{c|}{47} & 53 & \multicolumn{1}{c|}{46} & 54 \\ \hline
\textbf{Garcia-Munoz}       & \multicolumn{1}{c|}{49} & 51 & \multicolumn{1}{c|}{47} & 53 \\ \hline
\textbf{Maurin}             & \multicolumn{1}{c|}{49} & 51 & \multicolumn{1}{c|}{46} & 54 \\ \hline
\end{tabular}
\caption{Analysis of contained zeros from difference between semblance using transport equation solution convection-diffusion and force field and different LIS}
\label{analysisZeros}
\end{table*}

\newpage 
\hypertarget{conclusions}{%
\section{\texorpdfstring{\textbf{Conclusions}}{Conclusions}}\label{conclusions}}

Both analyses allow quantifying the correlation between the modulation factor and sunspots and anti-correlation between the modulation factor and the mean magnetic field. The models are mostly similar with both transport equation solutions and the local interstellar spectrum as they only show slight changes. 

For the first analysis, the the modulation factor yields similar outcomes and are solely dependent on the rigidity factor. 
The correlation between sunspots and modulation factor is more pronounced for stations closer to the magnetic poles, such as OULU and SNAE, than for the equatorial AATB station. However, this pattern is not replicated for the semblance between the modulation factor and the mean magnetic field. 
The station near the magnetic equator has shows more anti-correlation than near the poles. The pattern is similar for every semblance, even when different LIS and transport equation solutions were used.

Thanks to the second analysis, the first analysis can be affirmed and generalized, the force field solution is also more compatible with neutron detectors data. Additionally, any modulation factor solution will be similar to the neutron detector data and will only have minimal amplitude variations.

In the second analysis, it is possible to observe similarity among the graphs of contained zeros in the difference of semblances, particularly when the modulation factors have been computed by different LIS. However, if semblance is computed using the Garcia-Munoz, Mason and Simpson LIS in 1975 more differences are observed. These differences are visible in the analysis of contained zeros for semblance between averaged modulation factor and mean magnetic field.

According to the table  \ref{analysisZeros} semblance using sunspots and averaged modulation factors data is very similar, and the percentage of contained zeros is comparable when using the Webber \& Lockwood and Burguer and Potgieter LIS in 2000 or Garcia-Munoz and Ghelfi, Barao, Derome and Maurin LIS in 2017 are used. 

Therefore, the LIS, that have the best adjust to the semblance model are Lagner, Potgieter and Webber LIS in 2003, Burguer and Ghelfi, Barao, Derome and Maurin LIS in 2017 using force field solution as they offer more consistent values, as shown in Table \ref{analysisZeros}.

With both proposed models, it can be asserted that there is a correlation between modulation phenomena and sunspots, because the solar activity is less allowed to detect more rays from the Sun. The anti-correlation between modulation phenomena and the mean magnetic field is due to high solar activity. Therefore, fewer solar rays affect the arrival of cosmic rays from other sources.

Thanks to this proposed quantitative model, a more precise predictive model of cosmic rays with solar activity can be computed.

\clearpage
\bibliography{sample631}{}

\begin{thebibliography}{}
\expandafter\ifx\csname natexlab\endcsname\relax\def\natexlab#1{#1}\fi
\providecommand{\url}[1]{\href{#1}{#1}}
\providecommand{\dodoi}[1]{doi:~\href{http://doi.org/#1}{\nolinkurl{#1}}}
\providecommand{\doeprint}[1]{\href{http://ascl.net/#1}{\nolinkurl{http://ascl.net/#1}}}
\providecommand{\doarXiv}[1]{\href{https://arxiv.org/abs/#1}{\nolinkurl{https://arxiv.org/abs/#1}}}

\bibitem[{{Auger} {et~al.}(1939){Auger}, {Ehrenfest}, {Maze}, {Daudin}, \&
  {Fr{\'e}on}}]{Auger}
{Auger}, P., {Ehrenfest}, P., {Maze}, R., {Daudin}, J., \& {Fr{\'e}on}, R.~A.
  1939, Reviews of Modern Physics, 11, 288, \dodoi{10.1103/RevModPhys.11.288}

\bibitem[{{Bieber}(1999)}]{Bibber}
{Bieber}, J. 1999, in International Cosmic Ray Conference, Vol.~7, 26th
  International Cosmic Ray Conference (ICRC26), Volume 7, 61

\bibitem[{{Boschini} {et~al.}(2017){Boschini}, {Della Torre}, {Gervasi},
  {Grandi}, {J{\'o}hannesson}, {Kachelriess}, {La Vacca}, {Masi}, {Moskalenko},
  {Orlando}, {Ostapchenko}, {Pensotti}, {Porter}, {Quadrani}, {Rancoita},
  {Rozza}, \& {Tacconi}}]{DellaTorre}
{Boschini}, M.~J., {Della Torre}, S., {Gervasi}, M., {et~al.} 2017, \apj, 840,
  115, \dodoi{10.3847/1538-4357/aa6e4f}

\bibitem[{Burger {et~al.}(2000)Burger, Potgieter, \& Heber}]{Burguer}
Burger, R.~A., Potgieter, M.~S., \& Heber, B. 2000, Journal of Geophysical
  Research: Space Physics, 105, 27447,
  \dodoi{https://doi.org/10.1029/2000JA000153}

\bibitem[{Caballero-Lopez \& Moraal(2004)}]{Caballero}
Caballero-Lopez, R.~A., \& Moraal, H. 2004, Journal of Geophysical Research:
  Space Physics, 109, \dodoi{https://doi.org/10.1029/2003JA010098}

\bibitem[{Caballero-Lopez \& Moraal(2012)}]{Caballero-Lopez}
---. 2012, Journal of Geophysical Research: Space Physics, 117,
  \dodoi{https://doi.org/10.1029/2012JA017794}

\bibitem[{Clem \& Dorman(2000)}]{Dorman}
Clem, J., \& Dorman, L. 2000, Space Science Reviews, 93,
  \dodoi{10.1023/A:1026508915269}

\bibitem[{{Cooper} \& {Cowan}(2008)}]{Cooper}
{Cooper}, G.~R.~J., \& {Cowan}, D.~R. 2008, Computers and Geosciences, 34, 95,
  \dodoi{10.1016/j.cageo.2007.03.009}

\bibitem[{Crank \& Nicolson(1947)}]{crank_nicolson_1947}
Crank, J., \& Nicolson, P. 1947, Mathematical Proceedings of the Cambridge
  Philosophical Society, 43, 50–67, \dodoi{10.1017/S0305004100023197}

\bibitem[{Ferrari {et~al.}(1996)Ferrari, Ranft, Roesler, \&
  Sala}]{Ferrari_1996}
Ferrari, A., Ranft, J., Roesler, S., \& Sala, P.~R. 1996, Zeitschrift für
  Physik C: Particles and Fields, 71, 75, \dodoi{10.1007/s002880050149}

\bibitem[{{Fisk} {et~al.}(1974){Fisk}, {Goldstein}, {Klimas}, \&
  {Sandri}}]{Fisk}
{Fisk}, L.~A., {Goldstein}, M.~L., {Klimas}, A.~J., \& {Sandri}, G. 1974, \apj,
  190, 417, \dodoi{10.1086/152893}

\bibitem[{{Garcia-Munoz} {et~al.}(1975){Garcia-Munoz}, {Mason}, \&
  {Simpson}}]{Garcia-Munoz}
{Garcia-Munoz}, M., {Mason}, G.~M., \& {Simpson}, J.~A. 1975, \apj, 202, 265,
  \dodoi{10.1086/153973}

\bibitem[{Ghelfi {et~al.}(2017)Ghelfi, Barao, Derome, \& Maurin}]{Ghelfi}
Ghelfi, A., Barao, F., Derome, L., \& Maurin, D. 2017, Astronomy {\&}
  Astrophysics, 605, C2, \dodoi{10.1051/0004-6361/201527852e}

\bibitem[{{Gleeson} \& {Axford}(1968)}]{Gleeson}
{Gleeson}, L.~J., \& {Axford}, W.~I. 1968, Canadian Journal of Physics
  Supplement, 46, 937, \dodoi{10.1139/p68-388}

\bibitem[{James(1998)}]{James:2296388}
James, F.~a. 1998.
\newblock \url{https://cds.cern.ch/record/2296388}

\bibitem[{Langner {et~al.}(2003)Langner, Potgieter, \&
  Webber}]{LangnerAndPotgieter}
Langner, U.~W., Potgieter, M.~S., \& Webber, W.~R. 2003, Journal of Geophysical
  Research: Space Physics, 108, \dodoi{https://doi.org/10.1029/2003JA009934}

\bibitem[{{Luo} {et~al.}(2015){Luo}, {Zhang}, {Potgieter}, {Feng}, \&
  {Pogorelov}}]{Luo}
{Luo}, X., {Zhang}, M., {Potgieter}, M., {Feng}, X., \& {Pogorelov}, N.~V.
  2015, \apj, 808, 82, \dodoi{10.1088/0004-637X/808/1/82}

\bibitem[{Moskalenko {et~al.}(2002)Moskalenko, Strong, Ormes, \&
  Potgieter}]{Moskalenko_2002}
Moskalenko, I.~V., Strong, A.~W., Ormes, J.~F., \& Potgieter, M.~S. 2002, The
  Astrophysical Journal, 565, 280, \dodoi{10.1086/324402}

\bibitem[{Neidell \& Taner(1971)}]{Neidell}
Neidell, N.~S., \& Taner, M.~T. 1971, GEOPHYSICS, 36, 482,
  \dodoi{10.1190/1.1440186}

\bibitem[{{O'Neill}(2006)}]{O'Neill}
{O'Neill}, P.~M. 2006, Advances in Space Research, 37, 1727,
  \dodoi{10.1016/j.asr.2005.02.001}

\bibitem[{{Parker}(1965)}]{Parker}
{Parker}, E.~N. 1965, \planss, 13, 9, \dodoi{10.1016/0032-0633(65)90131-5}

\bibitem[{{Potgieter} {et~al.}(2012){Potgieter}, {De Simone}, {Vos}, {Boezio},
  \& {Di Felice}}]{VosII}
{Potgieter}, M., {De Simone}, N., {Vos}, E., {Boezio}, M., \& {Di Felice}, V.
  2012, in 39th COSPAR Scientific Assembly, Vol.~39, 1526

\bibitem[{{Putze} \& {Derome}(2014)}]{Putze}
{Putze}, A., \& {Derome}, L. 2014, Physics of the Dark Universe, 5, 29,
  \dodoi{10.1016/j.dark.2014.07.002}

\bibitem[{Sartini {et~al.}(2010)Sartini, Simeone, Pani, lo~bue, Marinaro,
  Grubich, Lobko, Etiope, Capone, \& Favali}]{Sartini}
Sartini, L., Simeone, F., Pani, P., {et~al.} 2010, Nuclear Instruments and
  Methods in Physics Research A

\bibitem[{Sekido \& Elliot(1985)}]{sekeido}
Sekido, Y., \& Elliot, H. 1985, Early history of cosmic ray studies: Personal
  reminiscences with old photographs.
\newblock \url{https://www.osti.gov/biblio/7169147}

\bibitem[{{Shikaze} {et~al.}(2007){Shikaze}, {Haino}, {Abe}, {Fuke}, {Hams},
  {Kim}, {Makida}, {Matsuda}, {Mitchell}, {Moiseev}, {Nishimura}, {Nozaki},
  {Orito}, {Ormes}, {Sanuki}, {Sasaki}, {Seo}, {Streitmatter}, {Suzuki},
  {Tanaka}, {Yamagami}, {Yamamoto}, {Yoshida}, \& {Yoshimura}}]{Shikaze}
{Shikaze}, Y., {Haino}, S., {Abe}, K., {et~al.} 2007, Astroparticle Physics,
  28, 154, \dodoi{10.1016/j.astropartphys.2007.05.001}

\bibitem[{Strong {et~al.}(2007)Strong, Moskalenko, \& Ptuskin}]{Strong_2007}
Strong, A.~W., Moskalenko, I.~V., \& Ptuskin, V.~S. 2007, Annual Review of
  Nuclear and Particle Science, 57, 285,
  \dodoi{10.1146/annurev.nucl.57.090506.123011}

\bibitem[{{Tan} \& {Ng}(1983)}]{TanandNg}
{Tan}, L.~C., \& {Ng}, L.~K. 1983, \apj, 269, 751, \dodoi{10.1086/161084}

\bibitem[{Taner \& Koehler(1969)}]{Taner}
Taner, M.~T., \& Koehler, F. 1969, GEOPHYSICS, 34, 859,
  \dodoi{10.1190/1.1440058}

\bibitem[{Teolis(2017)}]{Teolis}
Teolis, A. 2017, Computational Signal Processing with Wavelets,
  \dodoi{10.1007/978-3-319-65747-9}

\bibitem[{{Vos} \& {Potgieter}(2015)}]{Vos}
{Vos}, E.~E., \& {Potgieter}, M.~S. 2015, \apj, 815, 119,
  \dodoi{10.1088/0004-637X/815/2/119}

\end{thebibliography}
\bibliographystyle{aasjournal}

\clearpage
\hypertarget{Appendix}{%
\section{\texorpdfstring{\textbf{Appendix}}{Appendix}}\label{appendix}}

\begin{figure}
\begin{minipage}[c]{0.4\linewidth}
\centering
\includegraphics[width=1.0\textwidth]{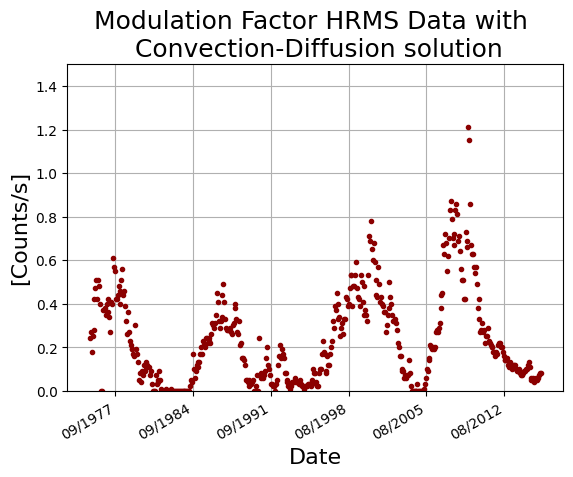}
    \caption{HRMS station. Modulation factor with Convection-Diffusion using Lagner, Potgieter \& Webber LIS in 2003}
    \label{Ap25}
\end{minipage}
\hfill
\begin{minipage}[c]{0.4\linewidth}
\centering
\includegraphics[width=1.0\textwidth]{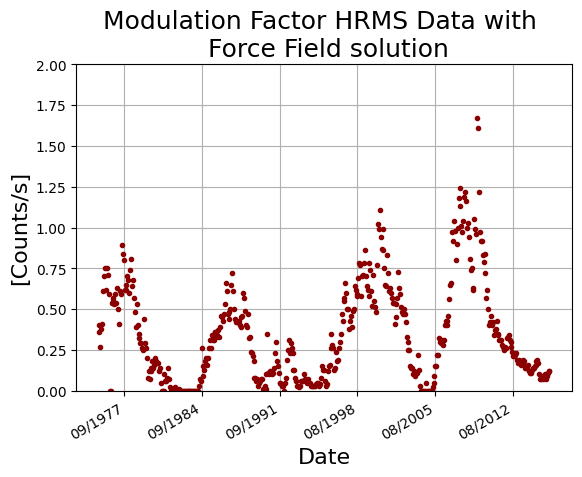}
    \caption{HRMS station. Modulation factor with Force Field using Lagner, Potgieter \& Webber LIS in 2003}
    \label{Ap26}
\end{minipage}%
\end{figure}

\begin{figure}
\begin{minipage}[c]{0.4\linewidth}
\centering
\includegraphics[width=1.0\textwidth]{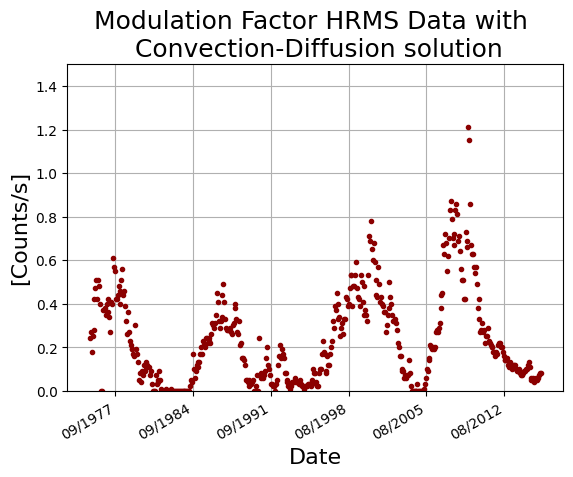}
    \caption{HRMS station. Modulation factor with Convection-Diffusion using Burguer and Potgieter LIS in 2000}
    \label{Ap27}
\end{minipage}
\hfill
\begin{minipage}[c]{0.4\linewidth}
\centering
\includegraphics[width=1.0\textwidth]{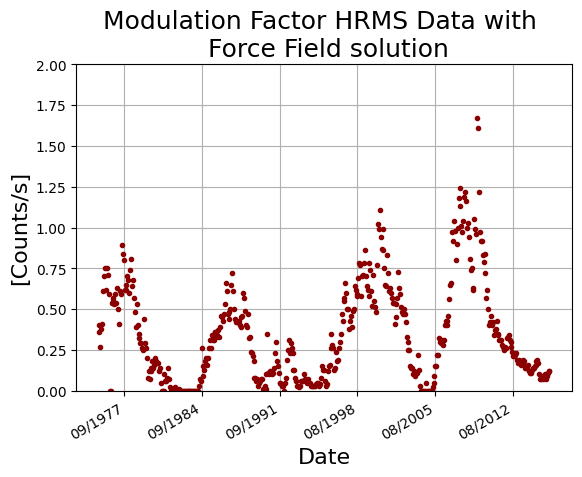}
    \caption{HRMS station. Modulation factor with Force Field using Burguer and Potgieter LIS in 2000}
    \label{Ap28}
\end{minipage}%
\end{figure}

\begin{figure}
\begin{minipage}[c]{0.4\linewidth}
\centering
\includegraphics[width=1.0\textwidth]{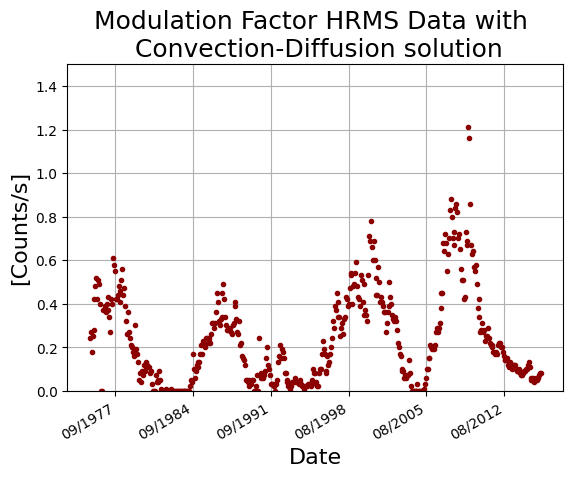}
    \caption{HRMS station. Modulation factor with Convection-Diffusion using Garcia-Munoz, Mason and Simpson LIS in 1975}
    \label{Ap29}
\end{minipage}
\hfill
\begin{minipage}[c]{0.4\linewidth}
\centering
\includegraphics[width=1.0\textwidth]{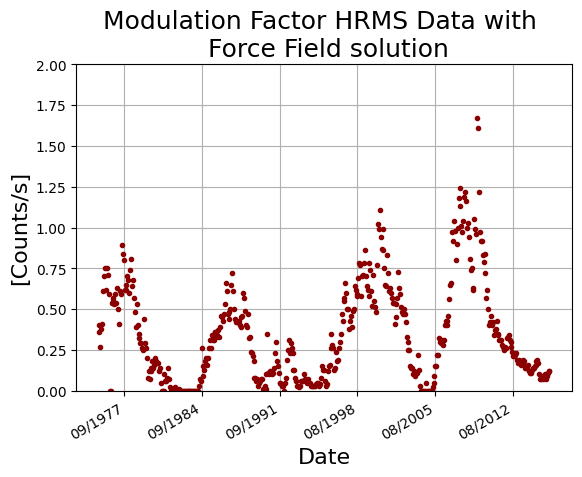}
    \caption{HRMS station. Modulation factor with Force Field using Garcia-Munoz, Mason and Simpson LIS in 1975}
    \label{Ap30}
\end{minipage}%
\end{figure}

\begin{figure}
\begin{minipage}[c]{0.4\linewidth}
\centering
\includegraphics[width=1.0\textwidth]{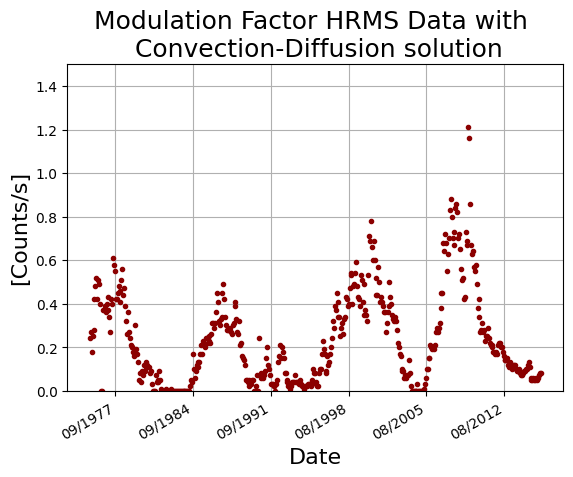}
    \caption{HRMS station. Modulation factor with Convection-Difussion using Ghelfi, Barao, Derome and Maurin LIS in 2017 in 2017}
    \label{Ap31}
\end{minipage}
\hfill
\begin{minipage}[c]{0.4\linewidth}
\centering
\includegraphics[width=1.0\textwidth]{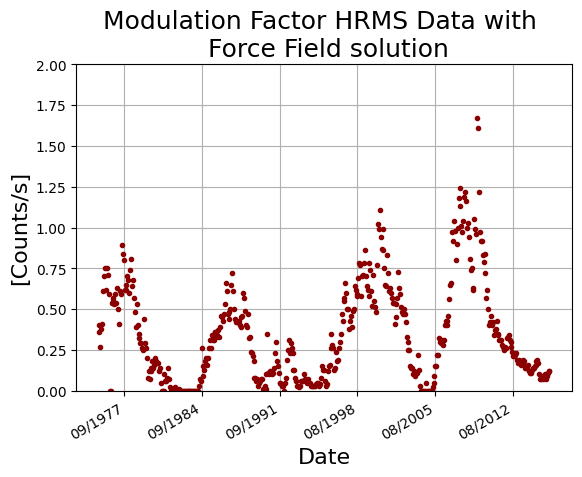}
    \caption{HRMS station. Modulation factor with Force Field using Ghelfi, Barao, Derome and Maurin LIS in 2017 in 2017}
    \label{Ap32}
\end{minipage}%
\end{figure}

\newpage
\begin{figure}
\begin{minipage}[c]{0.4\linewidth}
    \centering
    \includegraphics[width=1.0\textwidth]{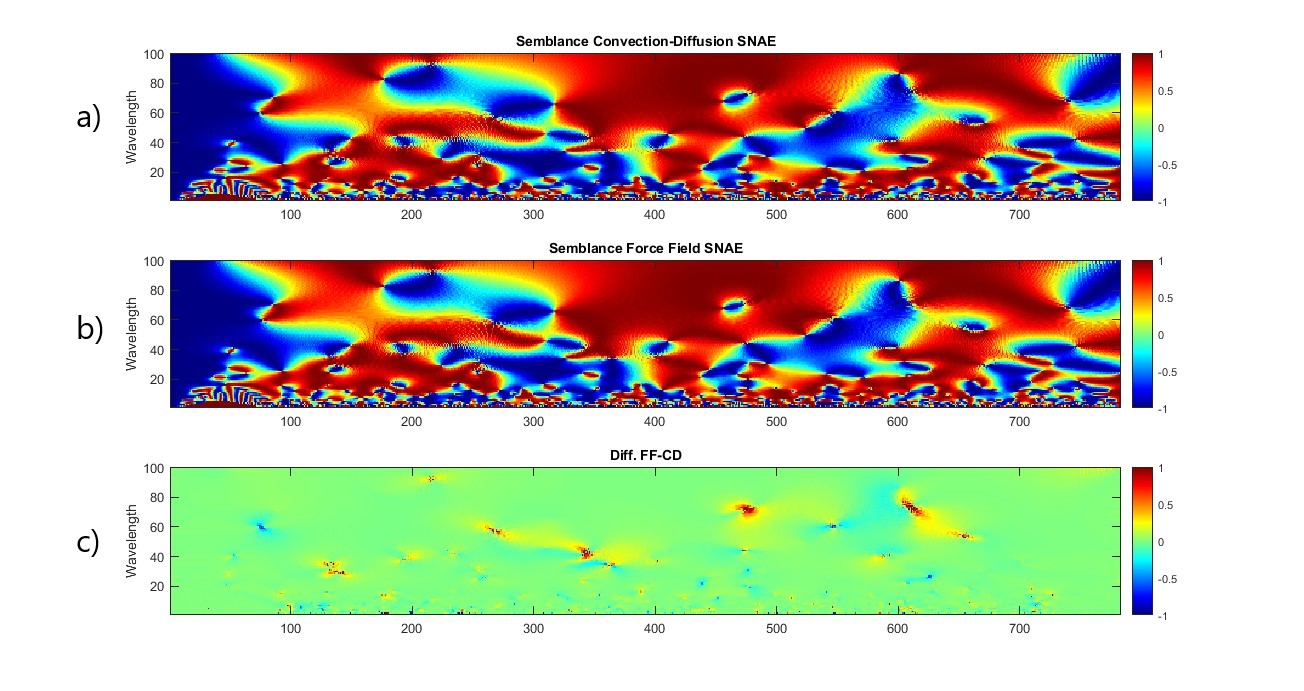} 
    \caption{SNAE Station a) Semblance between Modulation Factor using Convection-Diffusion with Lagner, Potgieter and Webber LIS in 2003 vs Sunspot data. b) Semblance between Modulation Factor using Force Field with Lagner, Potgieter and Webber LIS in 2003 vs Sunspot data. c) Difference between the two previous semblances.}
    \label{sem1}
\end{minipage}%
\hfill
\begin{minipage}[c]{0.4\linewidth}
    \centering
    \includegraphics[width=1.0\textwidth]{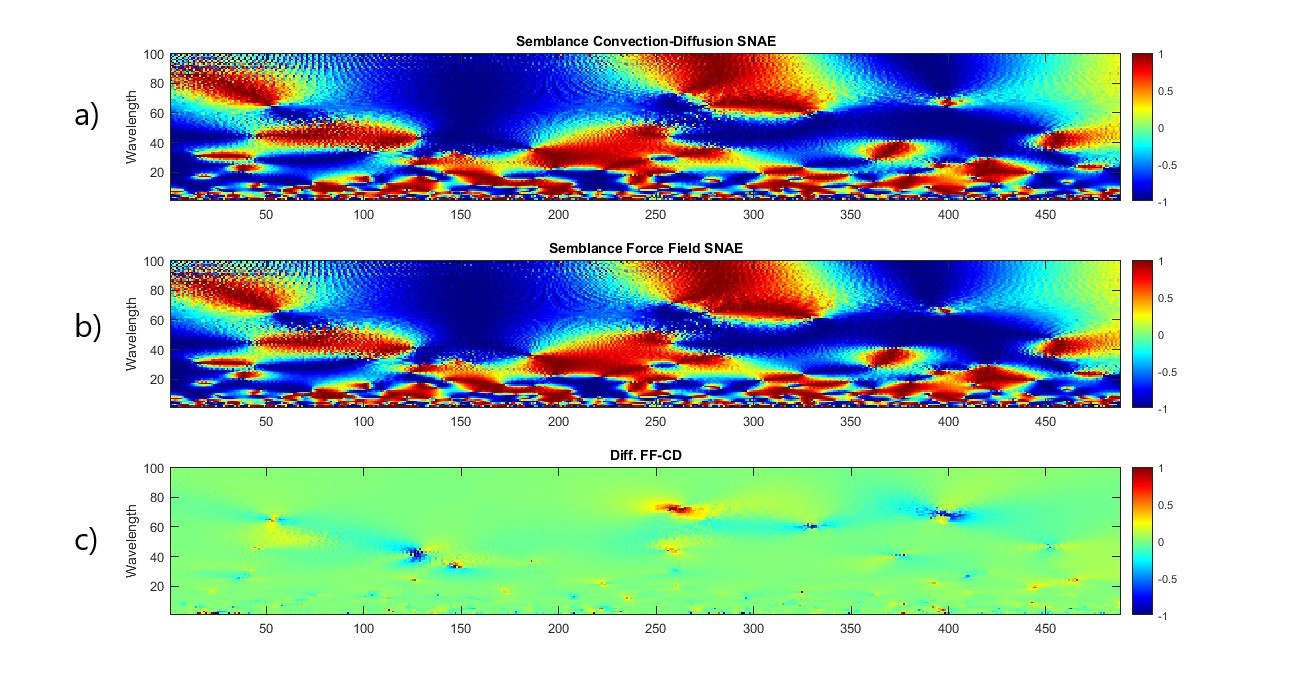}
    \caption{SNAE Station a) Semblance between Modulation Factor using Convection-Diffusion with Lagner, Potgieter and Webber LIS in 2003 vs Mean Magnetic Field data. b) Semblance between Modulation Factor using Force Field with Lagner, Potgieter and Webber LIS in 2003 vs Mean Magnetic Field data. c) Difference between the two previous semblances.}
    \label{sem4}
\end{minipage}%
\end{figure}

\begin{figure}
\begin{minipage}[c]{0.4\linewidth}
    \centering
    \includegraphics[width=1.0\textwidth]{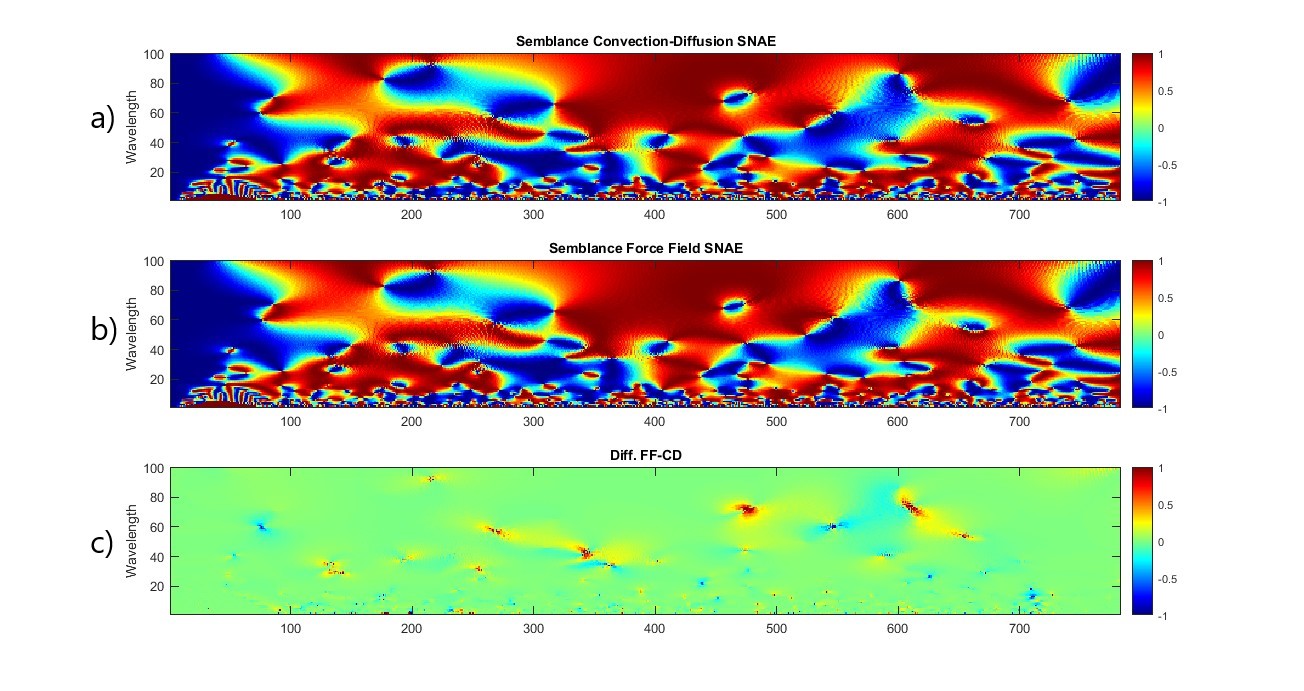}
    \caption{SNAE Station a) Semblance between Modulation Factor using Convection-Diffusion with Burguer and Potgieter LIS in 2000 vs Sunspot data. b) Semblance between Modulation Factor using Force Field with Burguer and Potgieter LIS in 2000 vs Sunspot data. c) Difference between the two previous semblances.}
    \label{sem7}
\end{minipage}
\hfill
\begin{minipage}[c]{0.4\linewidth}
    \centering
    \includegraphics[width=1.0\textwidth]{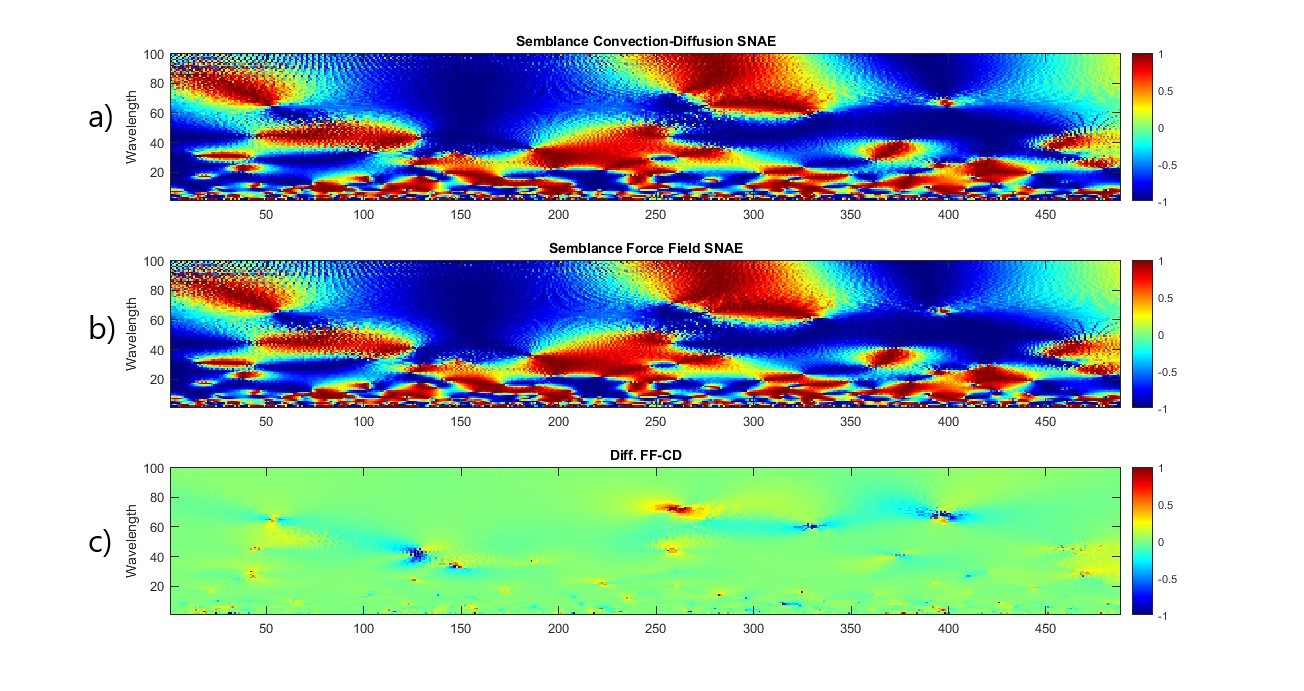}
    \caption{SNAE Station a) Semblance between Modulation Factor using Convection-Diffusion with Burguer and Potgieter LIS in 2000 vs Mean Magnetic Field data. b) Semblance between Modulation Factor using Force Field with Burguer and Potgieter LIS in 2000 vs Mean Magnetic Field data. c) Difference between the two previous semblances.}
    \label{sem10}
\end{minipage}
\end{figure}

\begin{figure}
\begin{minipage}[c]{0.4\linewidth}
    \centering
    \includegraphics[width=1.0\textwidth]{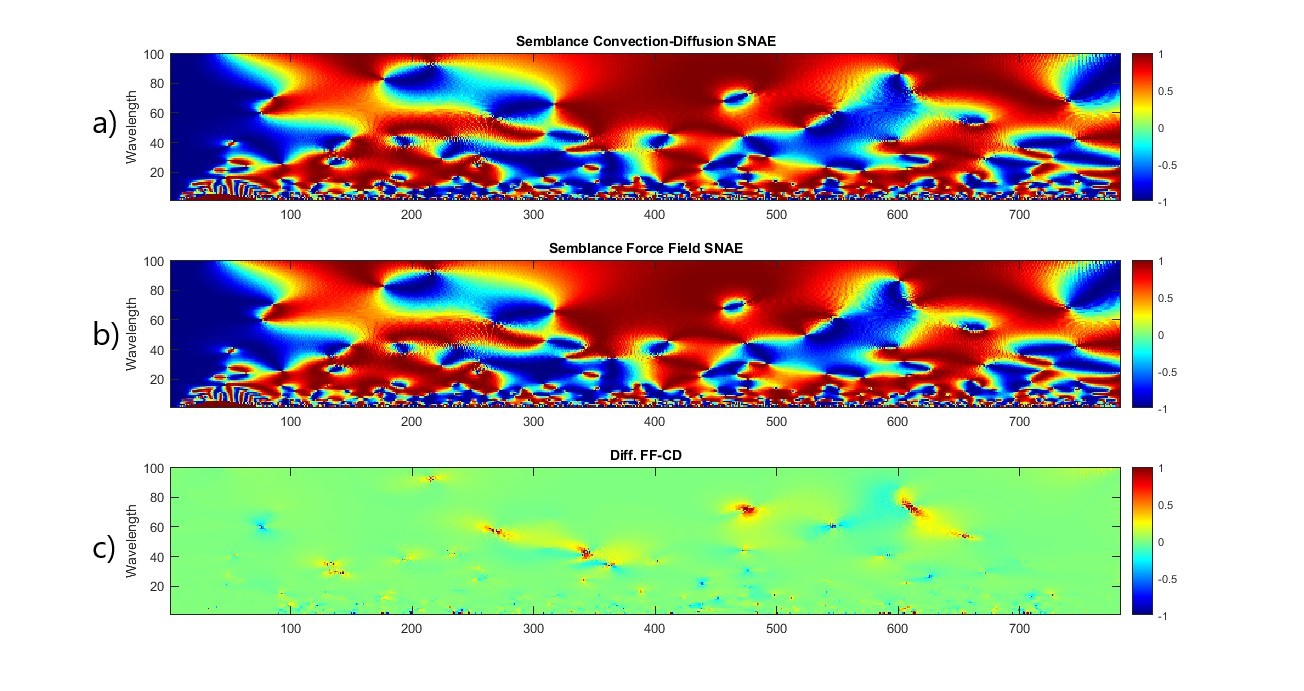}
    \caption{SNAE Station a) Semblance between Modulation Factor using Convection-Diffusion with Garcia-Munoz, Mason and Simpson LIS in 1975 vs Sunspot data. b) Semblance between Modulation Factor using Force Field with Garcia-Munoz, Mason and Simpson LIS in 1975 vs Sunspot data. c) Difference between the two previous semblances.}
    \label{sem13}
\end{minipage}
\hfill
\begin{minipage}[c]{0.4\linewidth}
    \centering
    \includegraphics[width=1.0\textwidth]{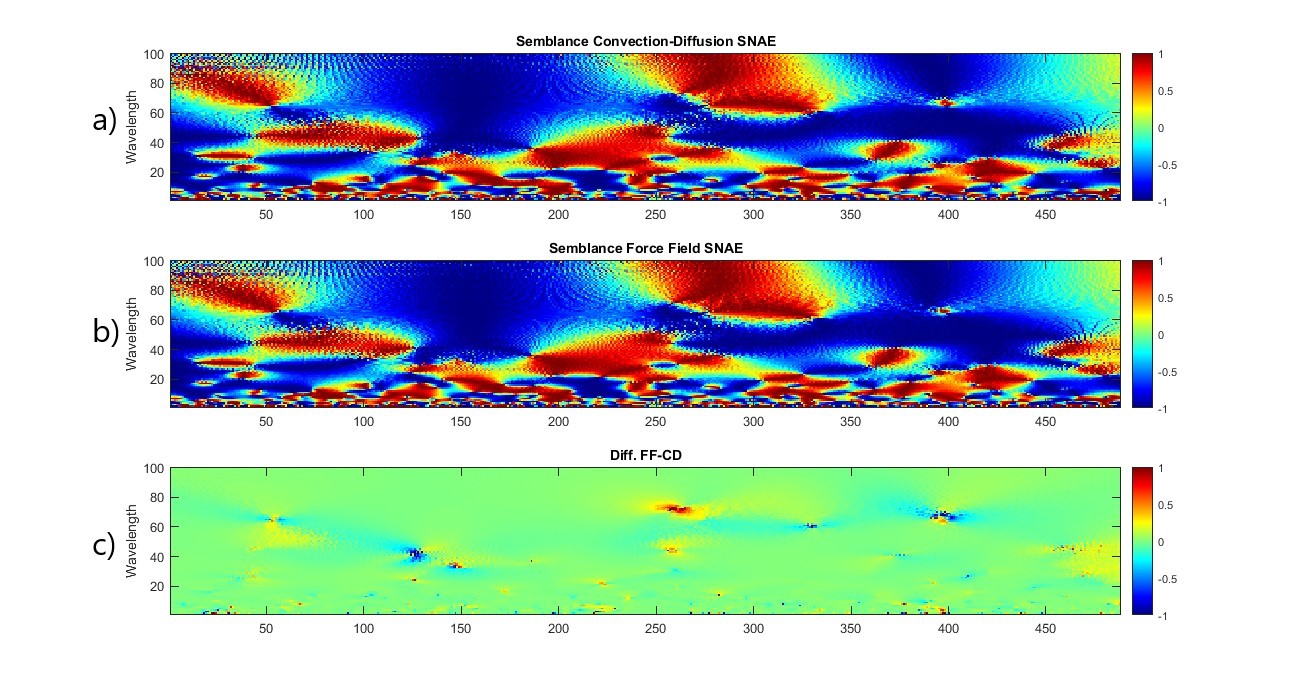}
    \caption{SNAE Station a) Semblance between Modulation Factor using Convection-Diffusion with Garcia-Munoz, Mason and Simpson LIS in 1975 vs Mean Magnetic Field data. b) Semblance between Modulation Factor using Force Field with Garcia-Munoz, Mason and Simpson LIS in 1975 vs Mean Magnetic Field data. c) Difference between the two previous semblances.}
    \label{sem16}
\end{minipage}
\end{figure}

\begin{figure}
\begin{minipage}[c]{0.4\linewidth}
    \centering
    \includegraphics[width=1.0\textwidth]{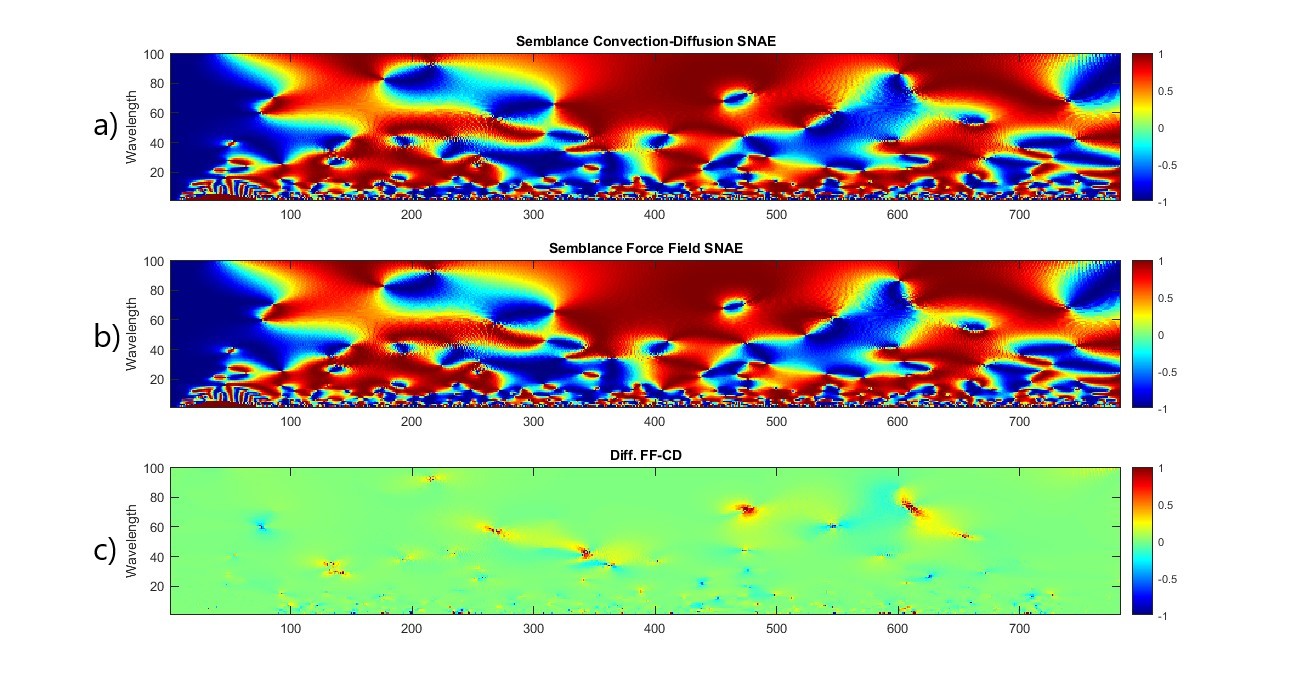}
    \caption{SNAE Station a) Semblance between Modulation Factor using Convection-Diffusion with Ghelfi, Barao, Derome and Maurin LIS in 2017 in 2017 vs Sunspots data. b) Semblance between Modulation Factor using Force Field with Ghelfi, Barao, Derome and Maurin LIS in 2017 in 2017 vs Sunspots data. c) Difference between the two previous semblances.}
    \label{sem19}
\end{minipage}
\hfill
\begin{minipage}[c]{0.4\linewidth}
     \centering
     \includegraphics[width=1.0\textwidth]{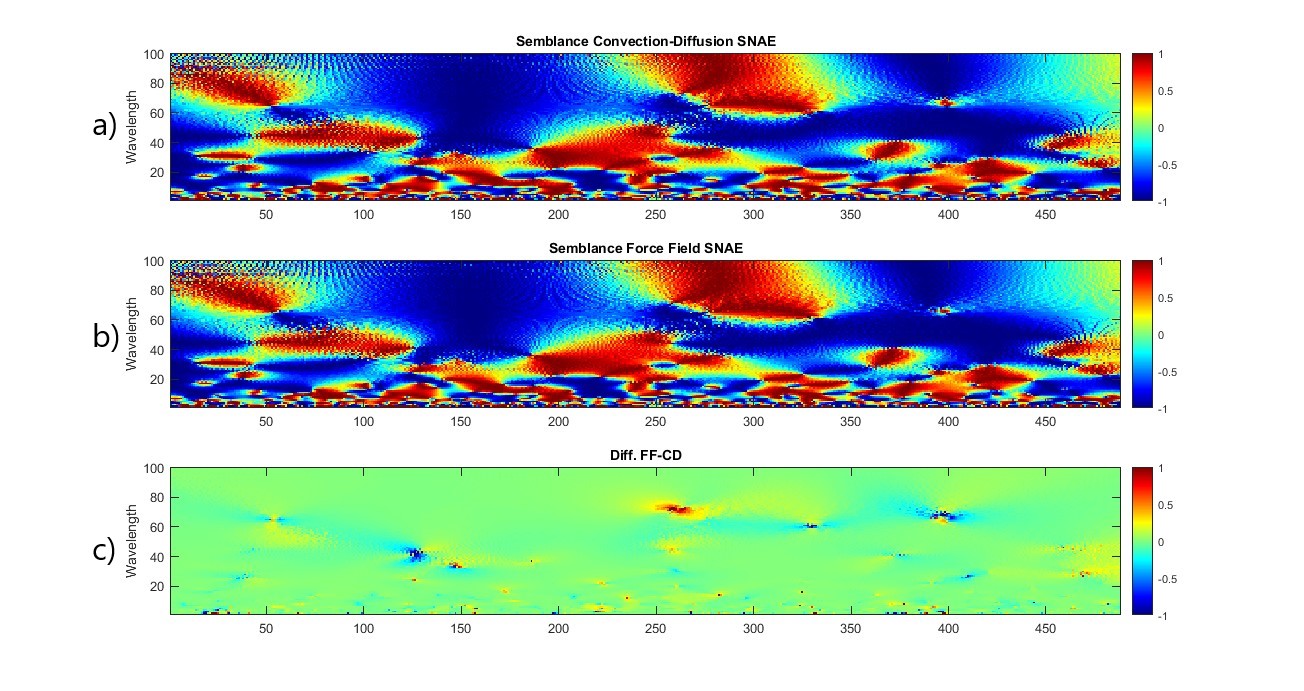}
     \caption{SNAE Station a) Semblance between Modulation Factor using Convection-Diffusion with Ghelfi, Barao, Derome and Maurin LIS in 2017 in 2017 vs Mean Magnetic Field data. b) Semblance between Modulation Factor using Force Field with Ghelfi, Barao, Derome and Maurin LIS in 2017 vs Mean Magnetic Field data. c) Difference between the two previous semblances.}
     \label{sem22}
\end{minipage}
\end{figure}
 


\end{document}